\documentclass[useAMS,usenatbib]{mn2e}
\usepackage{graphicx}
\usepackage{color}
\usepackage{aas_macros}

\def\be{\begin{equation}}
\def\ee{\end{equation}}
\def\kms{{\rm kms}}
\def\kpc{{\rm kpc}}
\def\Gyr{{\rm Gyr}}
\def\degree{{^o}}
\def\pc{{\rm pc}}

\def\mstream{{M_{\rm stream}}}
\def\Myr{{\rm Myr}}
\newcommand\msolar{{M_\odot}}

\title[A single-merger scenario for the formation of the giant stream and the warp of M31]
{A single-merger scenario for the formation of the giant stream and the warp of M31}
\author[Raphael Sadoun, Roya Mohayaee \& Jacques Colin]
{
Raphael Sadoun, Roya Mohayaee, Jacques Colin\\
CNRS, UPMC, Institut d'Astrophysique de Paris (IAP), 98 bis Boulevard Arago, Paris 75014, FRANCE
}

\date{\today}

\begin{document}
\maketitle
\label{firstpage}

\begin{abstract}

We propose  that the accretion of a dwarf spheroidal galaxy provides a common origin for the giant southern 
stream and the warp of M31.
We run about 40 full N-body simulations with live M31, infalling galaxies with varying masses and density profiles, and cosmologically-plausible initial orbital parameters. Excellent agreement with a full range of observational data 
is obtained for a model in which a dark-matter-rich dwarf spheroidal, whose trajectory lies on the thin plane of corotating satellites of M31, 
is accreted from its turnaround radius of about 200 kpcs into M31 at approximately  3 Gyrs ago. The satellite is disrupted as 
it orbits in the potential well of the galaxy and forms the giant stream and in return heats and warps the disk of M31. 
We show that our cosmologically-motivated model is favoured by the kinematic data over the phenomenological models in 
which the satellite starts its infall from a close distance of M31.
Our model predicts that the remnant of the disrupted satellite resides in the region of the North-Eastern shelf of M31.
The results here suggest that the surviving satellites of M31 that orbit on the same thin plane, as the disrupted satellite once did, could have all been 
accreted from an intergalactic filament. 

\end{abstract}


\begin{table*}
\centering
\begin{tabular}{lcccc}
\hline
\vspace{3pt}
Component & Model & Scale length(s) & Mass  & Additional parameters \\
          &       &   ($\kpc$)      & ($10^{10}\,\msolar$) &         \\
\hline
Disk & Exponential disk & Radial: $R_d = 5.40$ & 3.66 & \\
     &                  & Vertical: $z_0 = 0.60$ &  & \\
Bulge & Hernquist sphere & 0.61 & 3.24 & \\
Halo &  NFW sphere & 7.63 & $M_{200} = 88$ & $c = 25.5$ \\
     &             &      &    & $r_{200} = 195\,\kpc$ \\
\hline
\end{tabular}
\caption{Values of the parameters for different 
components of  M31, used in our simulations.}
\label{table:m31_model}
\end{table*}

\section{Introduction}
\label{sec:introduction}

A significant fraction of observed galaxies exhibit tidal features such 
as tidal tails, streams and shells \citep{malincarter1,malincarter2}. These features 
are widely believed to be the products of merger events 
\citep{hernquistquinn1,hernquistquinn2}. 
Numerous simulations have shown that tidal 
structures form during mergers of galaxies 
and observations of tidal structures have been used 
to put bound on various parameters, such as the orbital parameters 
and the masses of the host galaxies and their satellites.

In this work, we consider Andromeda or Messier 31 (M31) which is a rare example of a 
spiral galaxy that exhibits tidal features, such as streams and shells.
Andromeda galaxy contains two rings of star formation off-centered from the
nucleus (Block et al. 2006 and references therein) and most notably 
a Giant Southern Stream (GSS)
\citep{Ibata2001,Ferguson2002,Bellazzini2003,
Zucker2004,Ibata2005,Ibata2007,Richardson2008}. 
The giant southern stream is a faint stellar tail
located at the southeast part of M31. It extends radially outward
of the central region of M31 for approximately $5$ degrees, corresponding to a
projected radius of $\sim\,68\,\kpc$ on the sky. 
The stream luminosity is $ L_{\rm GSS} \sim 3.4\,\times\,10^7\,L_{\odot}$
corresponding to a stellar mass of 
$M_{\rm GSS} \sim 2.4\,\times\,10^8\,\msolar$ for a mass-to-light
ratio of $7$ \citep{Ibata2001,fardal2006}.

In the follow-up observations of the GSS, two other structures
corresponding to stellar overdensities, which are now believed to be
two shells, have been discovered \citep{Ferguson2002,fardal2007}. 
The colour-magnitude diagram of the North-Eastern shelf (NE) is
similar to that of the giant southern stream
\citep{Ferguson2005,Richardson2008}.
This similarity has been a strong argument in favour of models which predict
that both the GSS and the NE are the results of a single merger event between
M31 and a satellite galaxy \citep{Ibata2004,Font2006,fardal2007}.

A major merger scenario, dating back to a few Gyrs, from which M31, its giant stream, and many of its dwarf galaxies 
emerge has been proposed \citep{hammer2010,hammer2013}.
On the other hand, a phenomenological minor merger scenario has also been studied extensively, in which
a satellite galaxy falls onto M31, from a distance of a few tens of kpcs, on a highly radial orbit 
(of pericentre of a few kpcs) less than one 
billion year ago. The satellite is tidally disrupted
at the pericentre passage and forms the observed M31 stream and the two shelves 
\citep{fardal2006,fardal2007}. Although these empirical models 
provide good fits to the observations,
they suffer from simplifications. First, M31 is not modeled as a live galaxy but is only 
presented by a static potential and consequently the effect of 
dynamical friction is not properly taken into account.
Second, there is no dark matter in the progenitor satellite whereas a good fraction of satellite
galaxies in the local group seem to be dark-matter-rich. Finally, the origin
of the infalling satellite and its trajectory in the past is completely overlooked. It is highly implausible that
a satellite on a highly radial orbit could have survived to arrive at an easy reach of M31.

In the present work, we run full N-body simulations of mergers 
of satellites with a live M31. 
We take M31 as a live galaxy composing of a disk, a bulge and a dark matter halo 
of varying mass-to-light ratios and study the infalls of satellites 
with different density profiles, masses and orbital parameters.
Although a live realization of M31 has already been simulated for this model
to derive an upperlimit on the mass of the 
satellite (see {\it e.g.}\ \citet{Mori2008}), here we study
the dependence of the properties of the simulated stream 
on the internal structures of the progenitor and also study the history of the satellite itself.
First, we confirm that the empirical models, in which a dark-matter-poor satellite falls 
on a highly-radial orbit
from a short distance of a few tens of kiloparsecs, reproduce various observed
features of the giant stream of M31.
We study the orbital history of the satellite back 
in time and show that it is expected to have experienced several close 
encounters with M31  \citep{Ibata2004,Font2006}.
We demonstrate that a satellite that survives to reach 
within a short distance of its host halo is unlikely
to have followed a highly eccentric orbit.

We propose an alternative cosmologically-plausible scenario for the origin of the giant stream and also
the warp structure of M31 disk itself.
Here, a dark-matter-rich satellite is accreted and falls from its first turnaround radius, on an eccentric orbit onto M31.
The best agreement with the observational data is obtained when the satellite lies on
the same plane that contains 
many of the present dwarfs of M31 \citep{Ibata2013,conn2013}.
Unlike previous empirical models, the disk of M31 is perturbed by
the infall of the massive satellite in our model and becomes warped.

The paper is organised as follows. In Section 2, we present details 
of our numerical simulations and $N$-body modeling. 
In Section 3, we present results for the phenomenological models of GSS formation. Section 4 is devoted 
to the study of the orbital history of the satellite. In Section 5, we present 
the results for our alternative "first-infall" scenario. The perturbation, heating and warping of the disk of 
M31 due to the infall of the satellite are discussed in Section 6. We conclude in Section 7.

\section{Numerical methods}

\subsection{M31 mass model}
\label{section:M31}

The large spiral galaxy M31, at a distance of $d = 785 \pm 25$ kpc from Milky Way,
is probably the most massive, with a mass of 
$M_{300} = 1.4\pm 0.4\, \times\, 10^{12}\, \msolar$,  member of 
the local group.

The mass model of M31 that we use is based on previous works \citep{geehan2006}.
The disk of M31 is usually modeled with an exponential surface density profile which can be
written in cylindrical coordinates as : 
\be
\Sigma(R) = \Sigma_0 e^{-R/R_d} = \frac{M_d}{2\pi R_d^2}e^{-R/R_d}\,\,,
\ee
where $\Sigma_0$ is the disk central surface density, $R_d$ is the disk scale 
length in the radial direction
and $M_d$ is the mass of the disk. We set $R_d = 5.40\,\kpc$ 
and $M_d = 3.66\,\times\,10^{10}\,\msolar$ \citep{fardal2007}.
The disk has a finite thickness and its profile
in the vertical 
direction is assumed to be proportional to $\mathrm{sech}^2(z/z_0)$  
with a scale length $z_0 = 0.60\,\kpc$, which results in the density profile, 
\be
\rho(R,z) = \frac{\Sigma(R)}{2z_0}\mathrm{sech}^2(\frac{z}{z_0})\,.
\label{eq:expsechdisk}
\ee
The inclination and position angle of the disk are set to $77\degree$ 
and $37\degree$ respectively \citep{fardal2007}.

A spherical bulge modeled as a Hernquist profile \citep{hernquist1990} with a scale radius of
$r_b=0.61\,\kpc$ and a mass of $M_b=3.24\,\times\,10^{10}\,\msolar$ 
is also added to the model. 
The resulting density profile of the bulge is 
\be
\rho_b(r) = \frac{M_b}{2\pi}\frac{r_b}{r}\frac{1}{(r+r_b)^3}\,.
\label{eq:hern}
\ee

We also add a spherical dark matter halo with 
an NFW profile \citep{NFW1996} 
\be
\rho_h(r) = \rho_c\frac{\delta_c}{(r/r_h)(1+r/r_h)^2}\,, 
\label{eq:nfw}
\ee
where the parameter $r_h$ is the scale radius of the halo, $\rho_c$ is the background density
of the Universe at the current epoch and $\delta_c$ is the overdensity parameter. 
The concentration parameter $c$, which is the ratio of 
the scale radius to virial radius $r_{200}$, is set to $c=r_{200}/r_h=25.5$ and the mass within 
the virial radius is fixed at $M_{200} = 8.8\,\times\,10^{11}\,\msolar$.
The values of various structural parameters are given in Table \ref{table:m31_model}. 

To generate the N-body realization of M31,  we use the technique developed in previous works \citep{hernquist1993,springel1999}
which consists of approximating the velocity distribution 
by a 3-dimensional Gaussian whose moments are calculated using Jeans' equations. The halo of M31 is represented by
$N = 241369$ particles while the bulge and the disk have 
$N = 96247$ and $N = 108929$ respectively.
This ensures that the mass resolution for dark matter is, at most, 
ten times the mass resolution for the baryons, as given in  Table \ref{table:m31_num}.

\begin{table}
\centering
\begin{tabular}{ccc}
Component & $N$ & $m$ ($\msolar$) \\
\hline
Disk & $108929$ & $3.36\,\times\,10^5$\\
Bulge & $96247$ & $3.36\,\times\,10^5$ \\
Halo & $241369$ & $3.36\,\times\,10^6$ \\
\hline
\end{tabular}
\caption{Number of particles and mass resolution for our $N$-body realisation of M31.}
\label{table:m31_num}
\end{table}

\subsection{The satellite progenitor} 

\subsubsection{Morphology}

Based on the mass of the giant stream, which is found to be around $2.4\times 10^8\,\msolar$, 
and the extent of the giant stream, the stellar mass of the progenitor satellite 
has been estimated to be around $M = 2.2\,\times\,10^9\,\msolar$ \citep{Font2006,fardal2007}.
However, the morphology and the density profile of the progenitor are not immediately constrained 
by the giant stream and the shelves. 
Consequently, we have run simulations with different profiles and components.
In total we ran about 40 simulations, by varying the density profile, dark matter content and the initial
orbital parameters of the satellite. We group our simulations into two categories. 
The first category of the simulations uses a satellite with no dark matter 
and the common best-fit values of the orbital parameters \citep{fardal2007,fardal2013}.
We shall refer to these models as the empirical or phenomenological models. 
The simulation results for this category of models are presented in Section \ref{sec:empirical}.
In the second set of simulations, we search in different part of parameter 
space for models with a dark-matter-rich satellite and use cosmologically-motivated 
initial orbital parameters. The results corresponding to this category of models are presented 
and discussed in Section \ref{sec:infall}.

For each category of models, we run simulations with 
two different morphologies for the satellite: 
first we assume that the satellite
was a hot spheroid and run a simulation with
a Plummer profile of scale radius $a = 1.03\,\kpc$. It has already been reported 
that a satellite with such a profile satisfactorily reproduces 
the observed properties of the giant stream \citep{fardal2007,fardal2013}. The difference with
the previous works is that here we have a live M31 and 
consequently can properly take into account the effect of dynamical friction.
We shall refer to this as the Plummer model or shortly Plummer.
We also run two further simulations with spherical
Hernquist profiles, one with same scale radius $a = 1.03\,\kpc$ as the 
Plummer model and one with $a=0.55\,\kpc$, the later is chosen such that 
the half-mass radius of the Hernquist model is equal to that of the Plummer model. 
We shall refer to these models as Hernquist1 and Hernquist2.

In a second set of runs, we assume that the satellite was a cold 
rotating disk, which seems
to reproduce the {\it second-order} properties of the giant stream, in particular the 
observed asymmetry in the transverse density profile, even better than the previous examples of hot spheroids 
\citep{Mcco2003,Gilbert2007,fardal2008}.
We use a two-component model for the progenitor consisting of an exponential ($\mathrm{sech}^2$) 
disk as given by equation (\ref{eq:expsechdisk}), with a mass of $M_d=1.8\,\times\,10^9\,\msolar$, a 
scale radius of $R_d=0.8\,\kpc$ and a vertical scale length $z_d=0.4\,\kpc$, as well as a 
Hernquist bulge of mass $M_b=0.4\,\times\,10^9\msolar$ and scale radius $0.4\,\kpc$.
Because we lack constraint on the orientation of the disk, 
we consider six different models with evenly-spaced values of the inclination and 
position angles, $Ax$ and $Az$ respectively. We refer to 
these disk models as Disk$i$ with $i=1\cdots 6$.

\begin{table*}
\centering
\begin{tabular}{cccccc}
Model & Profile & Mass & Scale radius  & $Ax$  & $Az$  \\
      &         & ($10^9\,\msolar$) & ($\kpc$)  & ($\degree$) & ($\degree$) \\
\hline
Spherical & & & & & \\
\hline
Plummer & Plummer & $2.2$ & $1.03$ & & \\
Hernquist1 & Hernquist & $2.2$ & $1.03$ &          & \\
Hernquist2 & Hernquist & $2.2$ & 0.55 & & \\
\hline
Disk & & & & &  \\
\hline
Disk1 & Exponential & Disk = $1.8$ & $R_d = 0.8$ , $z_d = 0.4$ & 0 & 0 \\
   & Hernquist   & Bulge = $0.4$ & $r_b = 0.4$               &   & \\
Disk2 & \ldots & \ldots & \ldots & 45 & 0 \\
Disk3 & \ldots & \ldots & \ldots & 45 & 45 \\
Disk4 & \ldots & \ldots & \ldots & 45 & 90 \\
Disk5 & \ldots & \ldots & \ldots & 90 & 0 \\
Disk6 & \ldots & \ldots & \ldots & 90 & 45 \\
\hline
\end{tabular}
\caption{Values of parameters for different progenitor satellites 
used for an empirical modeling of the giant stream with live M31. The inclination angle $Ax$ and the position angle $Az$ are
of the satellite w.r.t. the disk of M31.}
\label{table:minorprofiles}
\end{table*}

\subsubsection{$N$-body realization: \textsc{NBODYGEN}}

The equilibrium $N$-body realizations of the progenitor satellites is 
generated by our code, \textsc{NBODYGEN}, which is specially tailored 
for the Plummer, Hernquist1 and Hernquist2 models. 

\textsc{NBODYGEN} is a code used to generate $N$-body realisations of multi-component elliptical and 
spheroidal galaxies with an optional central black hole and is available publicly at \citet{nbodygen}. 
The positions of particles for each component (bulge and halo) are selected 
by sampling the cumulative mass profile. The velocities are 
sampled from the self-consistent distribution function given 
by Eddington's formula (see {\it e.g.}~\citet{BT87,Kaz2004}).
The integrand in Eddington's formula is tabulated on a grid uniformly spaced in $x = r/(r+r_s)$ 
where $r_s$ is a characteristic radius of the profile. 
The distribution function is then calculated numerically on a grid of relative energy $\epsilon$ and linear interpolation 
is used whenever needed to obtain values other than the tabulated ones.

For the spherical Plummer and Hernquist profiles,
we run our simulations with a total number of $N = 131072$ particles to model the progenitor satellite. The disk progenitors are 
initialized using the same method as that used in the previous subsection to make 
the $N$-body realization of M31. In 
the case of disk models, the number of particles
in the disk is set to $N = 107143$ and in the bulge to $N = 23809$ in 
order to have the same particle mass resolution 
in both components. Given the chosen values for the number of particles 
and the progenitor mass, the particle mass 
in all models is $m_s = 1.68\,\times\,10^4\,\msolar$ (Table \ref{table:m31_num}).
The softening length is set to $\varepsilon = 30\,\pc$ 
for the satellite while it is 
$\varepsilon = 39\,\pc$ for the baryonic component and $\varepsilon = 390\,\pc$ for the 
halo of M31. 

\section{Empirical modeling of the M31 giant stream}
\label{sec:empirical}

\subsection{The orbital parameters}

Velocity and position measurements along the giant stream
have been used to constrain the orbital parameters of 
the progenitor satellite. In the first part of this study, we adapt 
the initial conditions \citep{fardal2007} :
\be
\begin{array}{lcl}
x_0 = -34.75 &,& v_{x0} = 67.34 \\
y_0 = 19.37  &,& v_{y0} = -26.12 \\
z_0 = -13.99 &,& v_{z0} = 13.50\;,
\label{eq:orbit_fardal}
\end{array}
\ee
where 
the positions are in $\kpc$ and the velocities in km/s. 
These best-fit parameters are calculated by fitting the orbital trajectory of the satellite to the observed
position and velocity data along the stream. 
In addition to the observed GSS data, the position of the 
NE shelf ($\xi=1.8\degree$,$\eta=0.7\degree$) 
was also used to constraint the initial orbital parameters. Various other 
similar models that find the orbits in slightly different potential 
have also been proposed \citep{Ibata2004,Font2006}. All of these models 
constrain the orbit of the progenitor to be highly radial.

\subsection{Identification of tidal structures}

\begin{figure*}
\centering
\includegraphics[width=\linewidth]{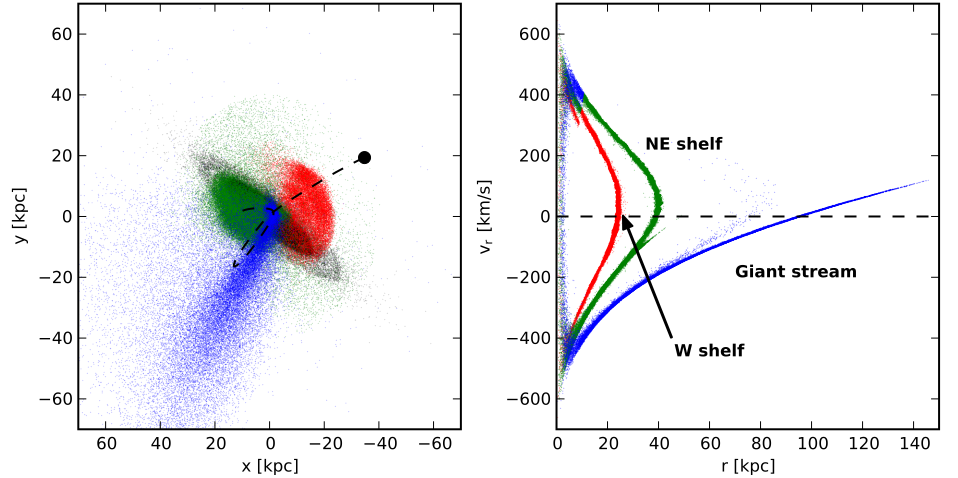} 
\caption{Real space $x-y$ (left panel) and phase space $r-v_r$ (right panel) projection of stellar particles 
in the progenitor satellite with a Plummer profile at the final time, $t = 0.84$ Gyr. On the $x-y$ projection, we also show
the particles that compose the disk of M31 (black dots) and the orbit of the progenitor (dashed line) as 
traced by the initially-most-tightly-bound particles. In both panels, the particles of the progenitor 
are colour-coded by the number of their pericentric passages: 1 (blue), 2 (green) or 3 (red).}
\label{fig:phaseplot_modelf}
\end{figure*}

The simulations are stopped at a time step that would yield the
best agreement between the simulated stream and
the observed stream and shelves, which corresponds to a time of $T\sim 840\,\Myr$.
Fig.~\ref{fig:phaseplot_modelf} shows the resulting real and phase-space 
projection for a satellite initialized with a Plummer profile. Particles are coloured 
by the numbers of pericentric passages that they experience during the run.  
We use the phase space plot to identify the shelves and the stream.
The Giant stream is easily identified and constitutes mostly
of stars with negative velocity with respect to M31. Its spatial extension 
can also be directly estimated from the phase-space plot and is $\sim140\,\kpc$
consistent with the observed value of $125\,\kpc$ \citep{Ibata2004}. 
The shelves manifest themselves as zero velocity surfaces in phase-space and 
hence are easily identified in a phase plot. Three of these phase structures
can be found in Fig.~\ref{fig:phaseplot_modelf}  and two of them are associated with the NE and W shelves.  
The third inner caustic has not yet been observed but is clearly a prediction of this model. 
The phase-space projection also clearly reveals that each tidal structure is formed 
of particles that went through equal numbers of pericentric passages. Thus, the tidal 
structures are formed by particles with similar initial 
orbital energies that have been stripped from the satellite.

\subsection{Spatial extent, morphology and stellar mass}

\begin{figure*}
\centering
\includegraphics[width=\linewidth]{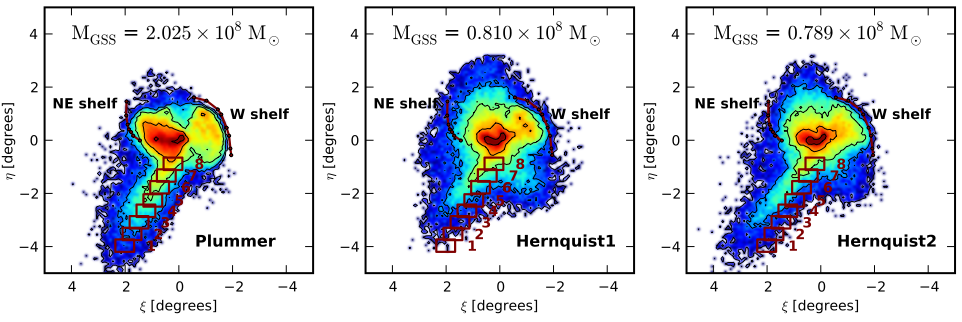}
\caption{
Spherical progenitors: Stellar density maps in standard sky coordinates 
corresponding to particles of the satellite for the 
three different spherical models; Plummer, Hernquist1 and Hernquist2 (from left to right).
The position of the observed stream fields from \citep{Mcco2003} are over plotted with the size of
the field-of-view of the CFH12k camera. The solid lines indicate 
the observed edges of the shelves. 
The total masses $\mathrm{M_{GSS}}$ of particles which are selected 
as stream members in each of our $N$-body models 
are indicated on each panel. 
Clearly, the satellites initialized with a cuspy Hernquist profile 
provide an overall poorer fit (for both position and stream mass) to the data than a core Plummer profile.
}
\label{fig:sky_spheremodels}
\end{figure*}

Next, we make a more detailed comparison with the observations. 
Fig.~\ref{fig:sky_spheremodels} shows the stellar density maps 
in standard sky coordinates at $T=840\,\Myr$ for the three models using a 
dynamically hot satellite with a spherical density profile, 
namely the fiducial Plummer model and the two Hernquist models (Hernquist1 and Hernquist2).
The fields covering the spatial region occupied by the GSS and  which have been used for 
the follow-up observations \citep{Mcco2003,Ibata2004} are plotted as solid rectangles 
with proper scaling on this figure.

For the Plummer model (left panel), we clearly see that the simulated stream is 
in good agreement with the observations regarding the morphology and spatial extent of the giant stream. 
The total mass $\mstream$ in the simulated stream can be calculated 
once the particles which did not originally belong to the satellite
have been removed. We find $\mstream = 2 \,\times\,10^8\,\msolar$ in excellent agreement 
with the value of $2.4\,\times\,10^8\msolar$ derived from observations. We can also compare
the spatial distribution of satellite particles with the positions of the edges of the shelves.
These edges are indicated as solid lines in Figure \ref{fig:sky_spheremodels} which 
are drawn by joining the data points \citep{fardal2007}.
We can see that there is still a fairly good agreement between the $N$-body model and the 
observation. In particular, the azimuthal and radial extent of the shelves are approximately 
reproduced with a better agreement for the Western shelf.

The Hernquist models do not succeed in reproducing correctly the proper apparent direction of the GSS on 
the sky. The deviations between the direction of the simulated and observed streams are not 
dramatic (a few degrees) but still indicate clearly that the Plummer model is a better fit to the data.
Furthermore, the total mass of the stream in the Hernquist1 and the Hernquist2 models
is a factor of $\sim2$ lower than the mass in the Plummer model. For these reasons, 
we only retain the Plummer model hereafter and shall refer to it simply as the 
spherical model.

\begin{figure*}
\centering
\includegraphics[width=0.8\linewidth]{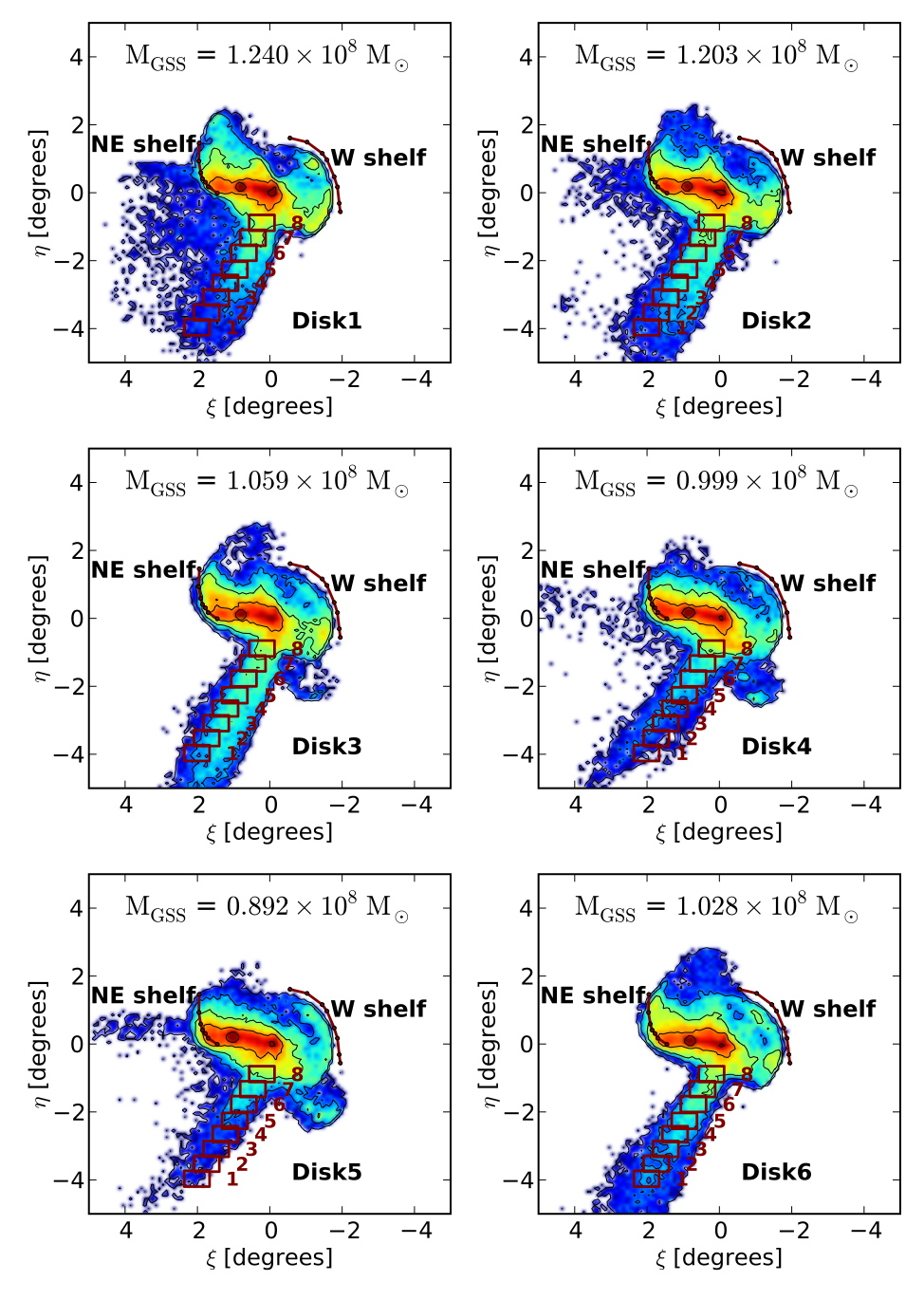}
\caption{Disk progenitors: Stellar density maps in standard sky coordinates at $t = 0.84$ Gyr for 
the six disk models of the satellite studied here. The six models differ only in the 
initial orientation of the disk of the infalling satellite w.r.t M31.
The plots have the same representation as the Figure \ref{fig:sky_spheremodels}. We see that all models, 
initialized with a disk progenitor, tend to underestimate 
the stream mass which has an observed value of $M_{\rm GSS} \sim 2.4\,\times\,10^8\,\msolar$. The best overall agreement 
with the GSS data (see also Fig.~\ref{fig:disk6_vs_data}) is obtained for model Disk6 corresponding 
to a satellite whose major axis is perpendicular to that of M31.
}
\label{fig:disk_sky}
\end{figure*}

Next we consider the six disky models for the satellite.
Fig.~\ref{fig:disk_sky} shows the stellar density maps in standard sky coordinates 
for each of the 6 Disk models which is to be compared to Fig.~\ref{fig:sky_spheremodels}. 
We recall that the only parameter that differs between these models is the initial 
inclination of the progenitor disk with respect to the M31 disk .
Let us first consider the spatial distribution of stream particles in each model. 
As can be seen from Figure \ref{fig:disk_sky}, the first three models (Disk1, Disk2 and Disk3) 
are able to reproduce the direction of the stream but substantially overestimate
its width. The remaining three models (Disk4, Disk5 and Disk6) consistently reproduce
the correct morphology of the stream with a slightly better agreement in the case of 
model Disk6. However, all models underestimate the total mass in the stream by a factor
of $~2$ similarly to the Hernquist spherical models that we have discussed previously. 
On the contrary, the shelves morphologies and spatial extent seem to be better 
reproduced by models Disk1, Disk2, Disk3
and Disk6 than by the Plummer model. The shelves 
in models Disk4 and Disk5 extend beyond the observed edges
indicated by the blue lines and moreover the azimuthal distribution is only poorly reproduced. 
Overall, we find that the disk progenitor that reproduces best both the morphology of 
the GSS and the shelves is Disk6 model 
which corresponds to the situation where the satellite disk's angular momentum is nearly aligned
with the major axis of M31. Consequently, hereafter, we only consider this model 
as a preferred disk model for the satellite.

\subsection{Distance and kinematics}

\begin{figure}
\centering
\includegraphics[width=\linewidth]{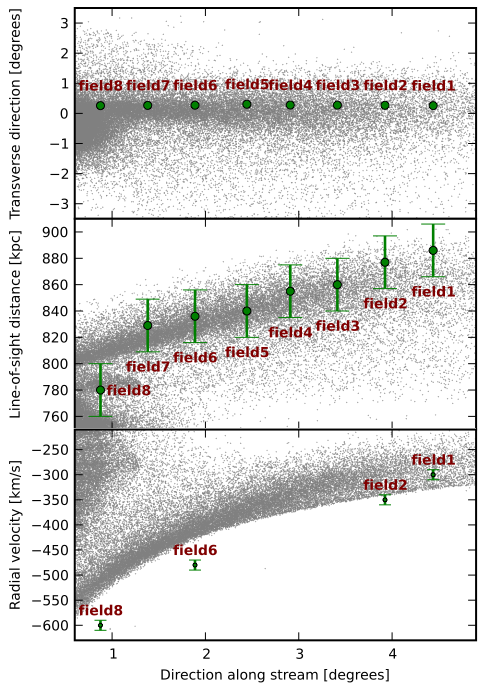} 
\caption{Comparison of $N$-body results from the Plummer model 
with positional and kinematical data of the GSS: 
position in stream-aligned coordinates (top panel), 
heliocentric distance (middle panel) and radial velocity (bottom panel) 
as a function of distance along the stream.
Green filled circles show the data points corresponding to the 
fields 1-8 of \citep{Mcco2003}. Radial velocity measurements are 
taken from \citep{Ibata2004} and are only available in four of these fields.
The particles of the progenitor in our simulation are represented as 
gray dots. The model is able to fit reasonably well the observations and 
can reproduce the distance-position correlation quite well. However, the phase plot (bottom panel) shows clearly that the 
velocity along the stream is mostly underestimated. 
}
\label{fig:modelf_vs_data}
\end{figure}

Next, we make a more in depth analysis of the spherical (Plummer) 
and the disk6 models by testing them against distance and kinematic data.
Fig.~\ref{fig:modelf_vs_data} shows the distribution of satellite particles (gray dots)
for the Plummer model together with the 8 fields of the position and velocity  observed data
\citep{Mcco2003,Ibata2004}. The upper panel corresponds to the projection in a sky coordinate 
system rotated such that the $x$ axis is aligned with the stream and the $y$ axis increases in the direction 
orthogonal to the stream. In this projection, the center of M31 is still located at the origin. 
We confirm that the morphology and spatial extent of the simulated stream agrees well with 
the position of the observed fields.

The middle and bottom panels show respectively 
the heliocentric distance and radial velocities as a function of distance along the stream. 
The observed line-of-sight distances are reproduced remarkably well by the $N$-body model 
which not only matches the observed values in individual fields but also 
the gradients along the axis of the stream. The only exception is for field 8 which is the 
nearest field from the center of M31. Therefore, it is likely that the distance estimate 
in this field is contaminated by M31 stars. The radial velocities, on the other hand, 
show larger discrepancy especially near the M31 disk. 
Apart from the first observational data farthest from M31, velocities along the stream are
systematically underestimated. 

\begin{figure}
\centering
\includegraphics[width=\linewidth]{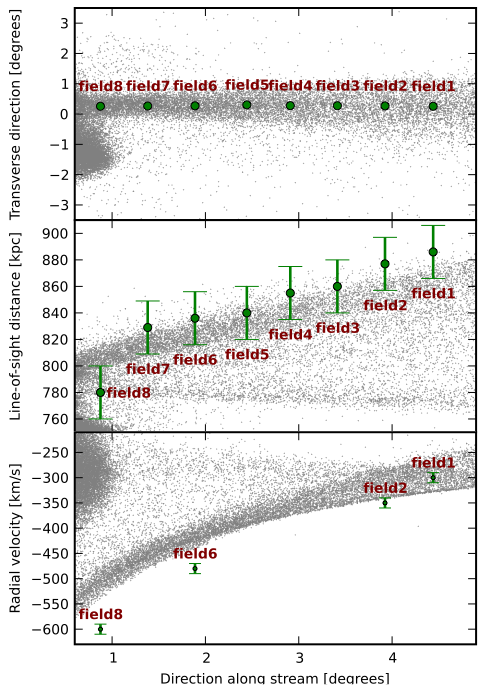}
\caption
{
The figure compares the results from our N-body simulation 
for model Disk6 with the observational data. This figure is similar 
to Fig.~\ref{fig:modelf_vs_data} but is now made for a disky satellite. Compared to the spherical Plummer 
satellite (Fig.~\ref{fig:modelf_vs_data}), the disk model shows a larger spread in 
the distance-position correlation (middle panel). It also fails to properly model
the radial velocity data which, apart from the farthest data point, are systematically underestimated (bottom panel).
}
\label{fig:disk6_vs_data}
\end{figure}

Fig.~\ref{fig:disk6_vs_data} shows the distribution of satellite particles
in model Disk6 at $T=840\,\Myr$ where the different panels refer to the same quantities as their correspondences  in 
Fig.~\ref{fig:modelf_vs_data}. The overall distribution of stream particles 
agrees with the position of the observed field. 
However, the line-of-sight distances are underestimated as compared to 
the observed values. This is to be compared to the spherical Plummer model, Fig.~\ref{fig:modelf_vs_data}, which produced a better fit to these data.
The Disk model, as the Plummer model, systematically underestimates the radial velocity along the stream, as shown in the bottom panel of 
Fig.~\ref{fig:disk6_vs_data}.

Overall, we find that, to first order,  there are no clear evidence to favor the disk model 
over the spherical Plummer model when including a live realization of M31. Both 
satellite morphologies are able to reproduce well the {\it first order} properties 
of the stream. 

\subsection{Number density profiles}

\begin{figure}
\centering
\includegraphics[width=\linewidth]{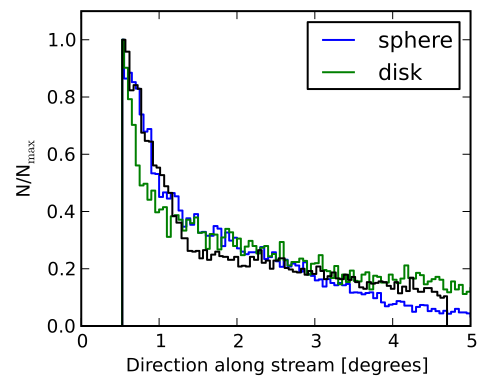}
\caption{
Stellar number density profile in the direction parallel to the stream. The black line 
shows the data  \citep{Mcco2003}. The blue line is the result from the spherical 
Plummer model while the green line corresponds to model Disk6. 
Since the number of stellar particles in the stream in our simulations is vastly superior to the 
number of observed stars, we normalize each profile by their respective maxima in order 
to be able to compare them directly to the observed profile. 
Furthermore, we exclude particles that are outside the region 
corresponding to the observed fields when calculating the profiles from our simulations. 
}
\label{fig:f_profile_along}
\end{figure}

\begin{figure}
\centering
\includegraphics[width=\linewidth]{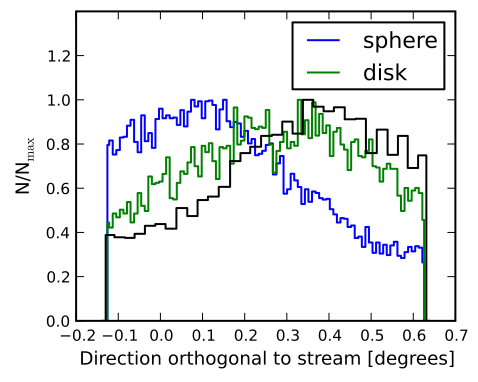}
\caption
{
Stellar number density profile of satellite stars in the direction orthogonal to the stream. 
The lines have the same meaning as in Figures \ref{fig:f_profile_along}. The GSS shows an 
asymmetry in the stellar distribution in the transverse direction
which is better reproduced by a cold disk satellite than a 
dynamically hot progenitor.
}
\label{fig:f_profile_ortho}
\end{figure}

Next, we test the disk and spherical models against 
{\it second-order} properties of the GSS. Fig.~\ref{fig:f_profile_along} and Fig.~\ref{fig:f_profile_ortho}
show the number density profile as a function of distance parallel and orthogonal to the stream 
respectively. 

Both models are able to reproduce the density profile along the stream but the 
spatial extension of the GSS ($\sim 5^o$) is better fitted by the spherical Plummer model.
Fig.~\ref{fig:f_profile_along} shows that both models produce a good approximative profile along the stream and hence
this data cannot be used to prefer one over another.
The observed profile in the transverse direction (Fig.~\ref{fig:f_profile_ortho}) is asymmetric with respect to the 
center of the observed fields with an excess in the north-eastern direction. 
This trend is captured correctly by the disk model which uses a rotationaly-supported 
satellite \citep{fardal2008}. On the other hand, the spherical Plummer model 
fails to reproduce this behaviour and, in fact, shows an excess in the south-west direction, contrary 
to the observational result.  The orthogonal profile of the stream is the only observation clearly in favour of a disky progenitor for the phenomenological models
whereas all other data seem to agree with both models almost equally well. 

\section{Where did the satellite come from ?}

\begin{figure}
\centering
\begin{tabular}{c}
\includegraphics[width=0.45\textwidth]{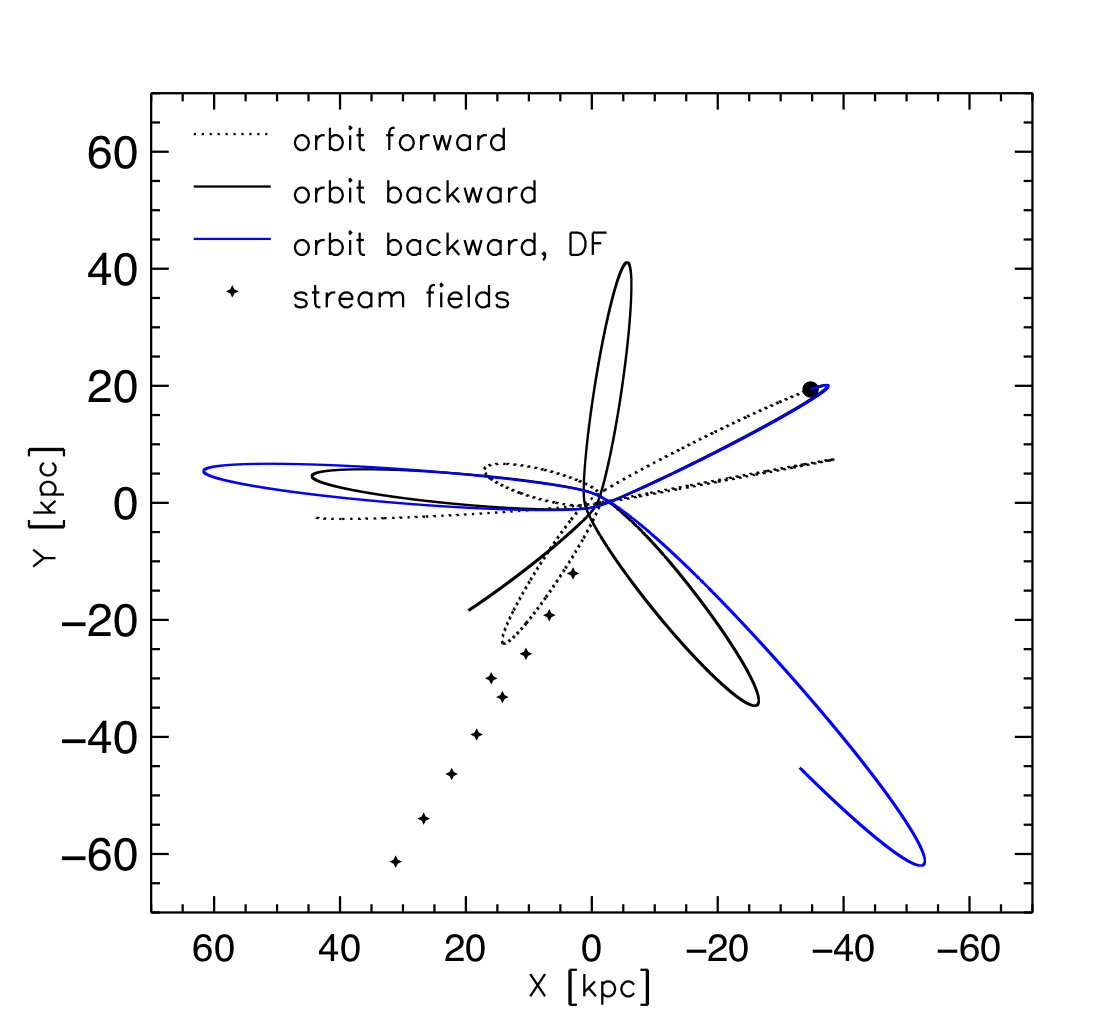} \\ 
\includegraphics[width=0.45\textwidth]{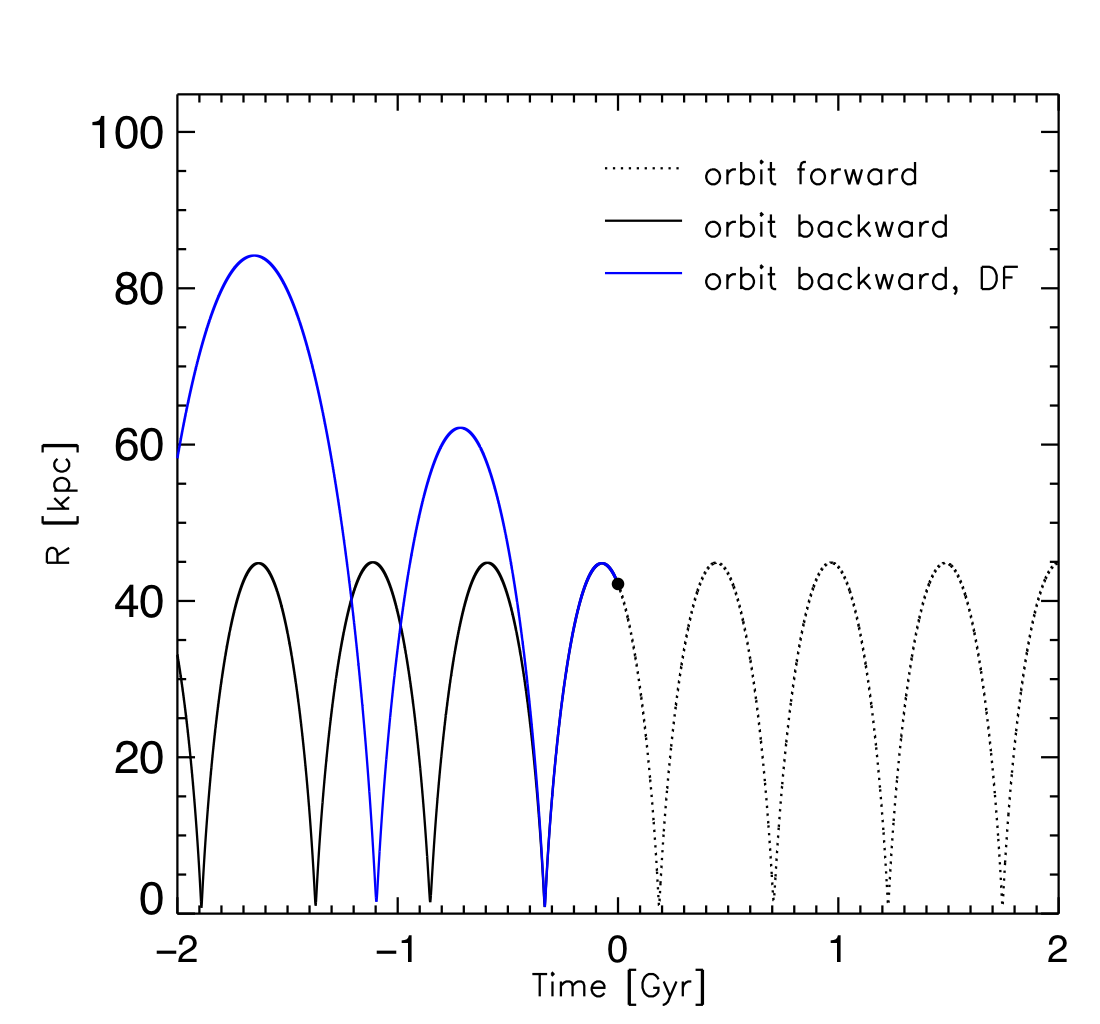} 
\end{tabular}
\caption{Tracing the progenitor orbit back in time: $(x,y)$ projection 
(top) and orbital radius evolution (bottom) for the 
numerically integrated orbits for initial conditions given by equation (\ref{eq:orbit_fardal}) for the empirical models. The dotted line represents the forward integration from the 
initial point (denoted with a filled circle) while the solid lines correspond to 
the past of the progenitor with (blue line) and 
without (black line) dynamical friction (DF).}
\label{fig:orbit_fiducial}
\end{figure}

So far in this work, we have used a single set of initial conditions,  
given by eq. ~\ref{eq:orbit_fardal}, for the progenitor satellite. 
This set of initial conditions assumes that the satellite started its infall onto
M31 around 800 Myrs ago at a separation of about 40 kpc. Although these models  
provide reasonable fits to the observations, they remain unmotivated and purely phenomenological.
Satellite galaxies are not usually accreted into their host haloes for the first time from a close random distance
and a satellite at a short distance from its host halo, on a highly radial orbit, is expected to 
have already been fully disrupted by the host on the previous pericentre passages.

If a satellite was at the origin of the GSS, then its origin itself need be properly traced back in time before a plausible set of initial conditions could be put forward.
Although most galaxies host satellites, the history
 of these satellites and their orbital characteristics remain obscure.
 Within the standard model of $\Lambda$CDM, a hierarchical formation of galaxies favours the  early formation of satellites and a later formation of their host galaxies.
 The initial motion of a satellite in its host potential is determined by the balance of two "forces": the Hubble expansion that pushes the satellite outwards and the gravitational potential of the host halo that
 pulls the satellite inwards.  The satellite, initially on Hubble flow, slows down under the 
 attraction of its host galaxy until its velocity is reduced to zero at which point it separates from the background Hubble expansion, turns around  and is accreted into the host galaxy. It is rather unlikely, that the satellite went through a turnaround and then arrived at (\ref{eq:orbit_fardal}),  as the present turnaround radius is 
by far larger than 40 kpc. (The present turnaround radius of M31 
is about 1 Mpc and the turnaround radius would roughly grow as $t^{8/9}$, given by a simple selfsimilar model \citep{FG1984,bert1985}) Satellites at such close separations from their host galaxies 
are most likely to have gone through a few orbits and 
to have arrived close to their host by losing energy through dynamical friction. 
However, a satellite 
that moves on a highly radial orbit would suffer 
disruption at its pericentre passages, such that it would not survive to reach a distance of 40 kpc from M31. 

To study this problem in details, we follow the trajectory of a particle in the potential of M31
back in time. Once again as for a live M31 in subsection \ref{section:M31}, we
model M31 as a Hernquist bulge with an NFW halo, but consider a Miyamoto-Nagai potential \citep{Miyamoto1975} for 
the disk in order to have an analytic expression for the potential.
In the backward integration, we also include a ``backward'' dynamical friction which is modeled using
the well-known Chandrasekhar's formula \citep{Chandra1943,BT87}. Clearly, we make the simplifying assumption
that the satellite loses little mass.
The orbit integration is performed using a leapfrog integrator,
because it is time-reversible and we can explore both forward and backward orbit integrations.
We begin our orbit integration from the initial condition (\ref{eq:orbit_fardal})
and integrate both forwards and backwards from this point
for up to $t=2\,\Gyr$ in either directions. For the backward integrations, we have considered
both cases with and without the dynamical friction. 

The resulting orbits are plotted in Figure \ref{fig:orbit_fiducial} which 
shows, in the top panel, the trajectory of the satellite in the $(x,y)$ plane, 
and in the bottom panel, the evolution of its orbital radius as a function of time. 
The plots clearly show that dynamical friction causes the test 
particle to gain energy when the orbit is integrated 
backwards. It is also clearly seen that the satellite has had several 
close pericentric passages ($r\sim 1\,\kpc$) prior to the GSS formation event. 
This is not entirely surprising since the characteristic of GSS constrains the satellite to be 
on a highly eccentric orbit. However, a dwarf galaxy such as the satellite
considered here is likely to be strongly disrupted after even a single one of these close encounters 
with M31.

A possible way out of this impasse would be to argue that the satellite
was a compact dwarf galaxy whose core survived repeated tidal shocks when passing through M31 \citep{Ibata2004}.
As a consequence, M32, a very dense satellite of M31, was suggested as a probable 
origin for GSS. However it was noted that 
the velocity and internal dispersion of M31 are difficult 
to match with the observed kinematics of the GSS and furthermore
M32 seems rather quiet and unperturbed.
It is worth mentioning that a collision between M32 and M31 has been investigated 
numerically in order to explain the ring structures observed 
at infrared wavelengths in the M31 disk \citep{Block2006}. 
Our various test simulations, have shown that
even the core of a satellite would not survive too many pericentre passages. 
This scenario would indeed require an unrealistically overdense galaxy to survive these passages and 
satisfy the observed properties of the giant stream.

A different and somehow far-fetched argument in favour of such a model is to assume that the satellite actually formed at a 
distance of around 40 kpc from M31 a few hundreds of megayears ago.
However, this is rather unlikely in a $\Lambda$CDM hierarchical model 
in which satellite galaxies are in general older and form earlier
than their parent galaxies and also it would be difficult to explain the stellar population of the 
giant stellar stream in such a model.

In the next section, we propose a new set of initial conditions that
are cosmologically-motivated and overcome the difficulties encountered in the 
empirical models for the formation of GSS 
and show that indeed such reasonably-conceived models do satisfy a full range of observational constraints.


\section{A dark-matter-rich satellite on its first infall}
\label{sec:infall}
\begin{figure*}
\centering
\includegraphics[width=0.45\textwidth]{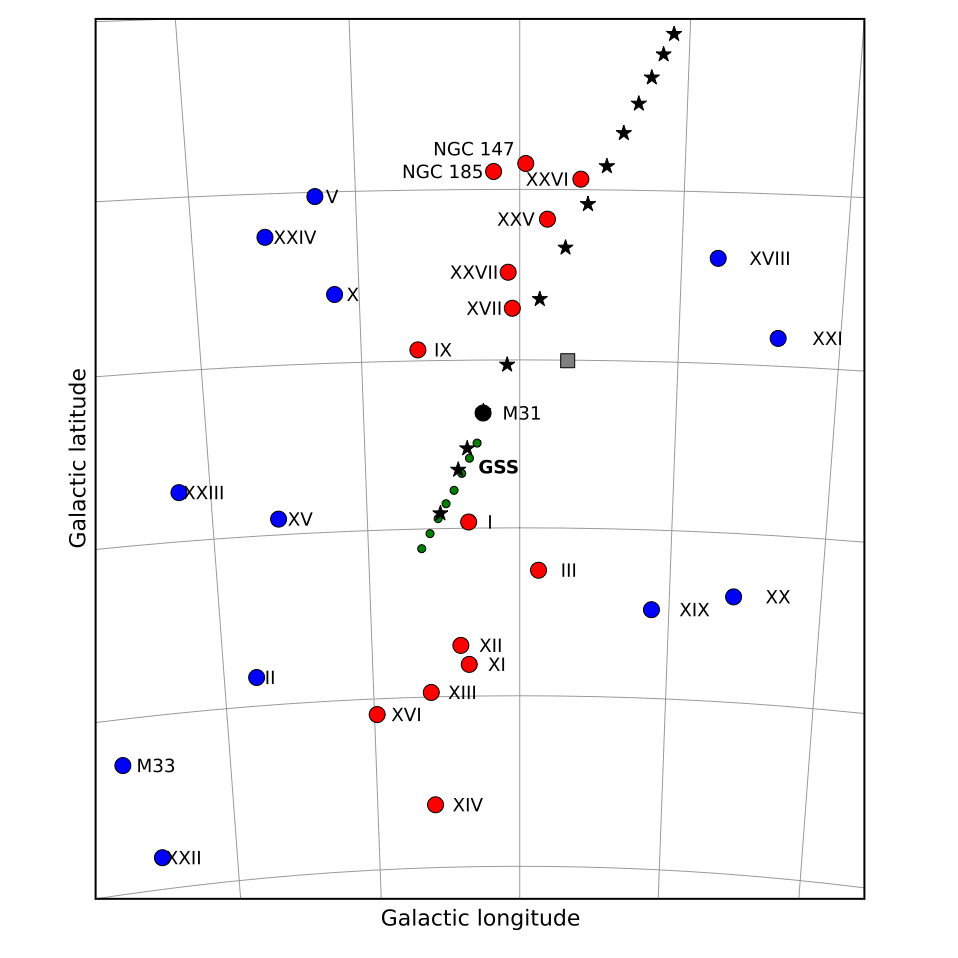} 
\caption{ 
The sky map of the satellites of M31 is shown (see \citet{Ibata2013,conn2013} for full details).
We have shown the positions of the giant stream (GSS) by the filled green circles. 
The grey box shows the initial position for 
the GSS progenitor in the previous empirical models (see equation (\ref{eq:orbit_fardal}) in Section \ref{sec:empirical}).
The orbit of the progenitor satellite given by the N-body simulation of our first-infall model  (see Section \ref{sec:infall}) is shown by stars, which lies almost on the same thin plane
as most of the satellites of M31.  In our model, the progenitor satellite is accreted from a large first turnaround distance of about 200 kpcs. 
}
\label{fig:thinplane-mw}
\end{figure*}
\begin{figure*}
\centering
\includegraphics[width=0.45\textwidth]{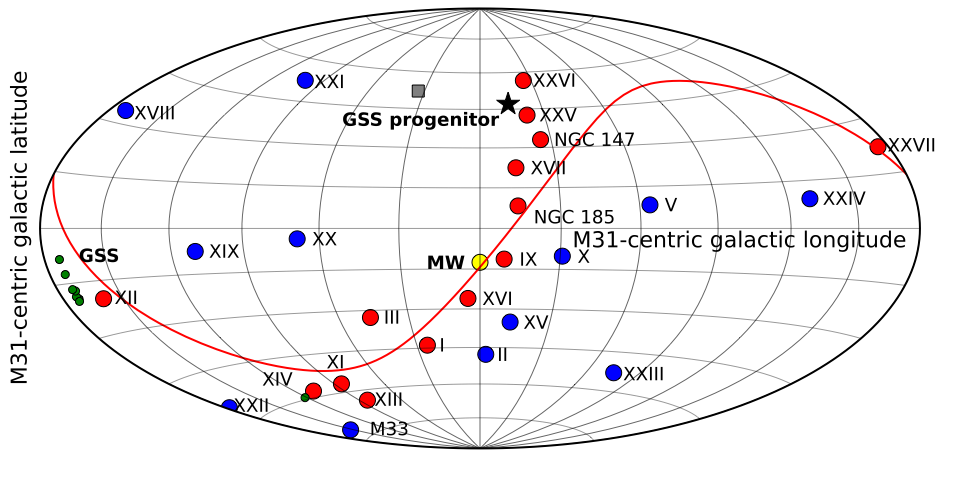}  
\caption{ 
This map is similar to Fig.~\ref{fig:thinplane-mw} but  all positions are now viewed from Andromeda.
We show the initial position of the satellite in our first-infall model  (see equation (\ref{eq:orbit_infall}) in Section \ref{sec:infall}) by star. 
The thin plane containing many of the M31 satellites is also drawn (see \citet{Ibata2013,conn2013} for full details).  The position
of the giant stream (GSS) is also shown on this map by green filled circles.}
\label{fig:thinplane-m31}
\end{figure*}

\subsection{Orbital parameters}

In a general cosmological set up, the accretion of a satellite, initially on Hubble flow,  into a galaxy occurs when it 
decelerates under the gravitational attraction of its host, reaches a zero velocity surface, turns around and
falls back into the host potential.
As the parameter space for our problem is unmanageably large,  we focus the initial conditions for our simulations around 
these cosmologically most-plausible configurations. In our model, the satellite starts its infall onto M31 from 
its first turnaround radius.  

An estimate of the turnaround radius of the satellite can be made as follows.
The turnaround radius grows, roughly, as
$
r_{\rm ta} \sim  t^{8/9}\;,
\label{eq:ta}
$
which is given by a simple secondary infall model \citep{FG1984,bert1985} for a highly radial and smooth accretion.
It has been shown that the secondary infall model represents quite well the numerical simulations in which dark matter haloes grow by
clumpy accretion of satellites \citep{ascasibar2007}.
It is hard to estimate the present turnaround radius of M31, as Milky Way and M31 are thought to have a common halo.
However, most observations put the present turnaround radius of M31 at around 1 Mpc.
Hence using the above expression, we see that a few Gyrs ago, the turnaround radius of M31 was about a few hundreds of kpcs.
We run around 40 full N-body simulations to fine-tune in this part of the parameter space.  
In our best-fit  model, the satellite starts at its first turnaround radius at about 200 kpcs with a null velocity and along 
the direction
\be
\begin{array}{lcl}
x_0 = -84.41\,\,, \\
y_0 = 152.47\,\,,  \\
z_0 = -97.08\,\,, \\
\label{eq:orbit_infall}
\end{array}
\ee
where the coordinates are given in kpc and in a reference frame centered on M31 with the $x$ axis pointing 
east, the $y$ pointing north and the $z$ axis correspond to the line-of-sight direction. These initial conditions 
have been found by sampling the parameter space in the region corresponding to the 
direction of the GSS observations, which is remarkably the same plane that is inhabited by the majority of 
the satellites of M31 \citep{Ibata2013,conn2013}, as shown in Fig.~\ref{fig:thinplane-mw} and Fig.~\ref{fig:thinplane-m31}.  

A satellite on such an orbit would have a very large velocity at the pericentre passage and would not be able to 
account for the large mass of the GSS, as it would loose too little mass. However, it could slow down by dynamical friction, which can
be significant if the satellite is dark-matter-rich. Therefore, we consider a dark-matter-dominated 
satellite which is also consistent with the observation of most local group dwarfs (see {\it e.g.}~ \citep{Mateo1998}). We assume that 
the stellar mass of the satellite is still the same as we used for the empirical models, {\it i.e.} $M_{s} = 2.2\,\times\,10^9\,\msolar$, studied in Section \ref{sec:empirical}.
This assumption is reasonable as the stellar mass of the satellite is relatively well-constrained by the stellar mass 
of the GSS \citep{fardal2006}. In our best fit initial conditions, the ratio of total to stellar mass 
is $M/M_{s} = 20$ and the halo has a mass of ${\rm M}_{\rm DM} = 4.18\,\times\,10^{10}\,\msolar$.

\subsection{Spatial extent, morphology and stellar mass}

In order to assess the ability of our model to reproduce the GSS observations, 
we perform a similar analysis as we did in Section \ref{sec:empirical}.
As before, the total time $T$ of the simulation is chosen such that to obtain a best match between our simulated 
stream and the GSS. With our cosmologically-motivated scenario, we find $T = 2.7\, \Gyr$s. Thus, the overall 
merger timescale in our scenario is much longer than for the phenomenological models which had $T = 0.84\,\Gyr$s.

\begin{figure*}
\centering
\includegraphics[width=\linewidth]{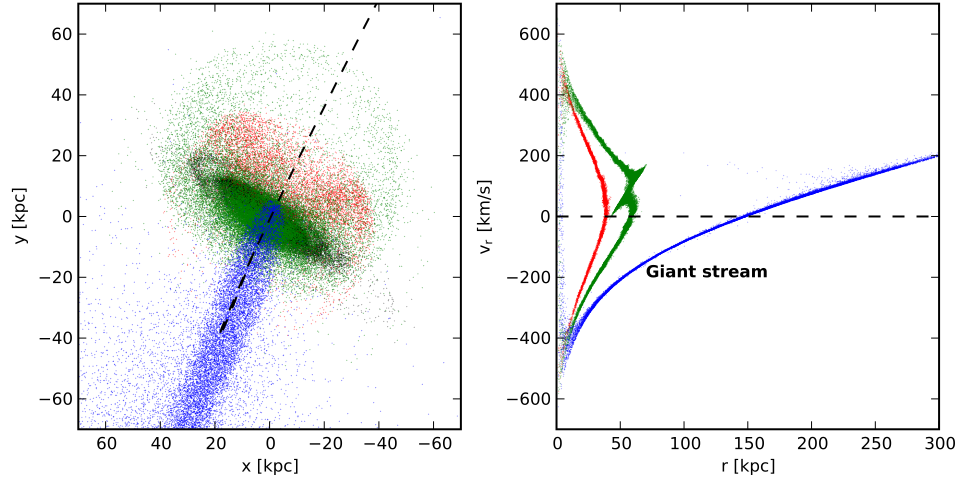}
\caption{
Distribution of stellar particles in the satellite in the $x-y$ 
plane (left panel) and the phase plot in the $r-v_r$ plane (right panel) at $t = 2.7$ Gyr for our first-infall 
$N$-body model using a spherical progenitor. The plots represent the same quantities as in 
Figure \ref{fig:phaseplot_modelf} but are now made for our first-infall model. Although difficult to identify in real space, the 
presence of two tidal caustics corresponding to a second and a third orbital wrap  
are clearly seen in the phase space (right panel).  The north-eastern shelf shown in green and the western shelf shown in red,  near the zero velocity surfaces, correspond to stars on their
second and third pericentre passages, respectively.
The position of the remnant of the satellite, which lies in the region of the north-eastern shelf, is also clearly seen in 
this plot which is in agreement with results from a recent statistical approach \citep{fardal2013}.
}
\label{fig:phaseplot_infall}
\end{figure*}

First, we test the spatial distribution, morphology and stellar mass in the stream. 
Fig.~\ref{fig:phaseplot_infall} shows the real space (left panel) and phase space (right panel)
projections of stellar particles initially in the satellite similarly to Fig.~\ref{fig:phaseplot_modelf}.
For clarity, the dark matter particles of both M31 and the satellite have been omitted from these plots.
We trace the orbit of the satellite by following the initially most-bound particles in our simulation. 
The resulting trajectory is represented as a dashed line in the real space 
projection and shows that the merger is almost a head-on collision between M31 and the satellite. 
We are also able to identify the tidal structures in phase-space and find again the presence 
of an extended stream and two caustics formed by coherent group of particles with same number of 
pericentric passages. 

\begin{figure*}
\centering
\includegraphics[width=\linewidth]{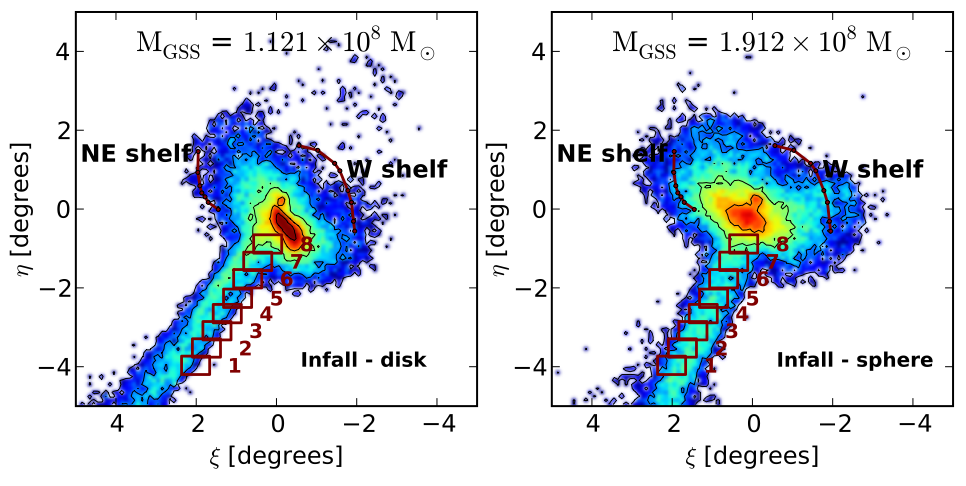}
\caption
{
Stellar density maps in standard sky coordinates similar to Figure \ref{fig:sky_spheremodels} but 
for our cosmologically motivated ``first infall'' model plotted for a cold disky progenitor (left panel) and a hot spheroidal dwarf progenitor (right panel). Our simulations favour a dynamically hot spheroidal dwarf  over a cold disky progenitor.
}
\label{fig:skymaps_infall}
\end{figure*}

The stellar density maps in sky coordinates 
compared to the position of the observed fields of the GSS and the edges of the two shelves is shown in Fig.~\ref{fig:skymaps_infall}.
We obtain a good agreement between our simulated stream and the GSS in the case of the spherical 
Plummer model. On the other hand, the disk model is unable to reproduce correctly the angular direction 
of the stream. We also calculate the stellar mass of the simulated stream and 
find $M_{\mathrm{GSS}} = 1.912\,\times\,10^{8}\,\msolar$ for the spherical model in good agreement with the estimated 
GSS mass and $M_{\mathrm{GSS}} = 1.121\,\times\,10^{8}\,\msolar$ for the disk model as indicated by Fig.~\ref{fig:skymaps_infall}.
This further confirms that the disk model provides a poorer fit to the GSS than a spherical satellite.

\subsection{Distance and kinematics}

\begin{figure}
\centering
\includegraphics[width=\linewidth]{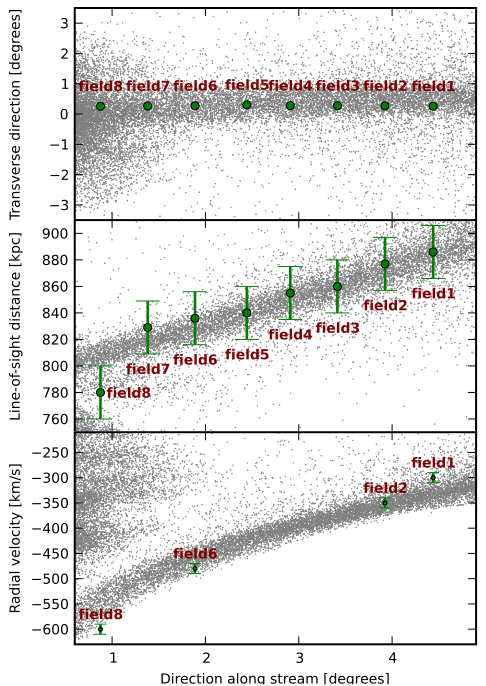}
\caption
{
Comparison of results from our $N$-body simulation of the first-infall model using a spherical 
Plummer satellite with observational data for the GSS.
The figure 
is similar to Fig.~\ref{fig:modelf_vs_data}. 
Our cosmologically-motivated first-infall scenario for the formation 
of the giant stream successfully reproduces the stream's three-dimensional position 
and kinematics.   
}
\label{fig:gss_comparison_infall_sphere}
\end{figure}

\begin{figure}
\centering
\includegraphics[width=\linewidth]{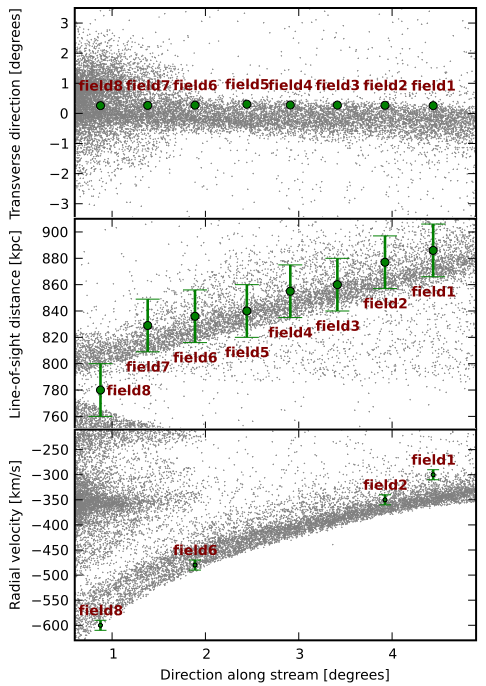}
\caption
{
Comparison of results from our $N$-body simulation of the first-infall model 
using a disky satellite (grey dots)
to the observational data (green points with error bars). The figure 
is similar to Fig.~\ref{fig:modelf_vs_data}. 
The agreement with the data is poorer than that of our model 
with a spherical progenitor (see Fig.~\ref{fig:gss_comparison_infall_sphere}). 
}
\label{fig:gss_comparison_infall_disk}
\end{figure}

Next, we examine the three dimensional distribution of the stream 
and its kinematics. Fig.~\ref{fig:gss_comparison_infall_sphere} and \ref{fig:gss_comparison_infall_disk}
show the comparison between our simulated stream and observations, similar to what we did in Fig.~\ref{fig:modelf_vs_data}
and \ref{fig:disk6_vs_data} for phenomenological models. We obtain an excellent agreement with observations for both the spatial distribution 
as well as the heliocentric distance and radial velocity measurements. The agreement is significantly better 
for the spherical model than for the disk model.
In particular, for the spherical model, the scatter in the distance-position correlation 
is fully consistent with the distance error estimates from \citet{Mcco2003} except 
for field 8 which is most-likely due to contamination 
from M31 disk stars since this field is the closest from M31 center. 

The kinematic data, from the observations of GSS, provides a strong evidence in support of our model. The 
phase plot, bottom panel in Fig.~\ref{fig:gss_comparison_infall_sphere} follows the motion of the satellite as it falls into and is disrupted by M31 and forms the giant stream.
Our simulations show that the kinematic data favours an infall from a large initial radius.
 Previous empirical models produce a less satisfactory agreement with 
 the velocity data and have a large velocity offset, because in these models, studied in Section \ref{sec:empirical}, the satellite starts its infall 
at a short distance of about 40 kpcs from M31 and hence the velocity of stream particles, throughout the orbit, are smaller than suggested by the observations. A short initial infall radius means a shorter subsequent apocentre and hence a smaller velocity along the trajectory.

\subsection{Number density profiles}

\begin{figure}
\centering
\includegraphics[width=\linewidth]{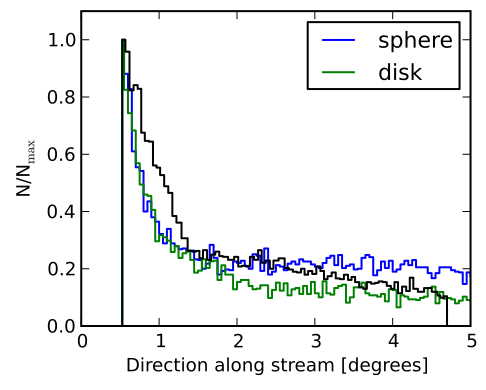}
\caption{
Density profile of satellite stars calculated in the direction parallel to the stream 
(same as done for Fig.~\ref{fig:f_profile_along}) for our first-infall models.
The black line 
shows the data  \citep{Mcco2003}. The blue line is the result from the spherical 
Plummer model while the green line corresponds to a disky satellite. 
We find minor deviations from the observed profile (black line) with a 
steeper shape at small distances and a shallower behavior at large radii for both 
a spherical (blue line) or disk (green line) satellite. The length 
of the stream is broadly consistent with its observed value. 
}
\label{fig:profile_along_infall}
\end{figure}

We also investigate the density profiles of the stream in our models. Similarly to the procedure described
in Section \ref{sec:empirical}, we only consider particles in the region defined by the observed fields. 
Fig.~\ref{fig:profile_along_infall} shows the density profile in the direction parallel to the stream. 
We find that, for both the spherical (blue line) and the disk (green line) models, the shape of the simulated profile
differs slightly from the observed one (black line). In our models, the density decreases more rapidly in the inner 
region of the stream but presents a shallower slope at large distances. 

\begin{figure}
\centering
\includegraphics[width=\linewidth]{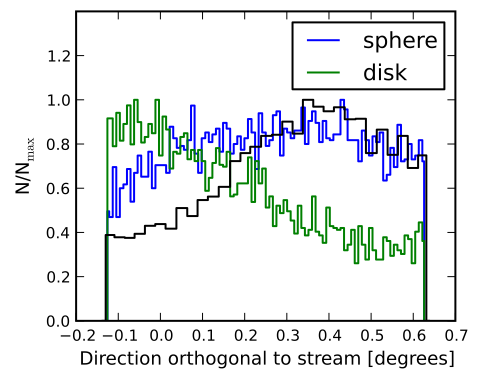}
\caption
{
Density profile of satellite stars calculated in the direction transverse to the stream 
as presented previously in Figure \ref{fig:f_profile_ortho} but now 
for our first-infall models. 
The black line 
shows the data  \citep{Mcco2003}. The blue line is the result from the spherical 
Plummer model while the green line corresponds to a disky satellite. 
The spherical progenitor clearly reproduces 
better the asymmetric shape of the profile than the disky satellite.  
}
\label{fig:profile_ortho_infall}
\end{figure}

The density profile in the transverse direction is shown in Fig.~\ref{fig:profile_ortho_infall}.
 For our first-infall scenario, the spherical model reproduces well 
the asymmetric profile orthogonal to the stream but the disk model fails to do so. 

\subsection{Velocity dispersion}

So far, we have compared the kinematics of the stream with the observed GSS kinematics 
using the radial velocity measurements given by \citet{Ibata2004}. However, we can further 
test the viability of our model  
by comparing the velocity dispersion at difference radii from the center of M31, to those given by the observations. 
From the radial velocity measurements in fields 1,2, 6 and 8, \citep{Ibata2004},
the mean observed velocity profile along the stream has been found and fitted by 
$v_h(\eta) = -4244.8\tan \eta - 610.9 \kms$, where $\eta$ is the North-South direction 
in standard sky coordinates. It is then possible to derive an estimate of the velocity 
dispersion along the stream as the offset between this mean profile and the velocity 
of each stellar particle.

\begin{figure}
\centering
\includegraphics[width=\linewidth]{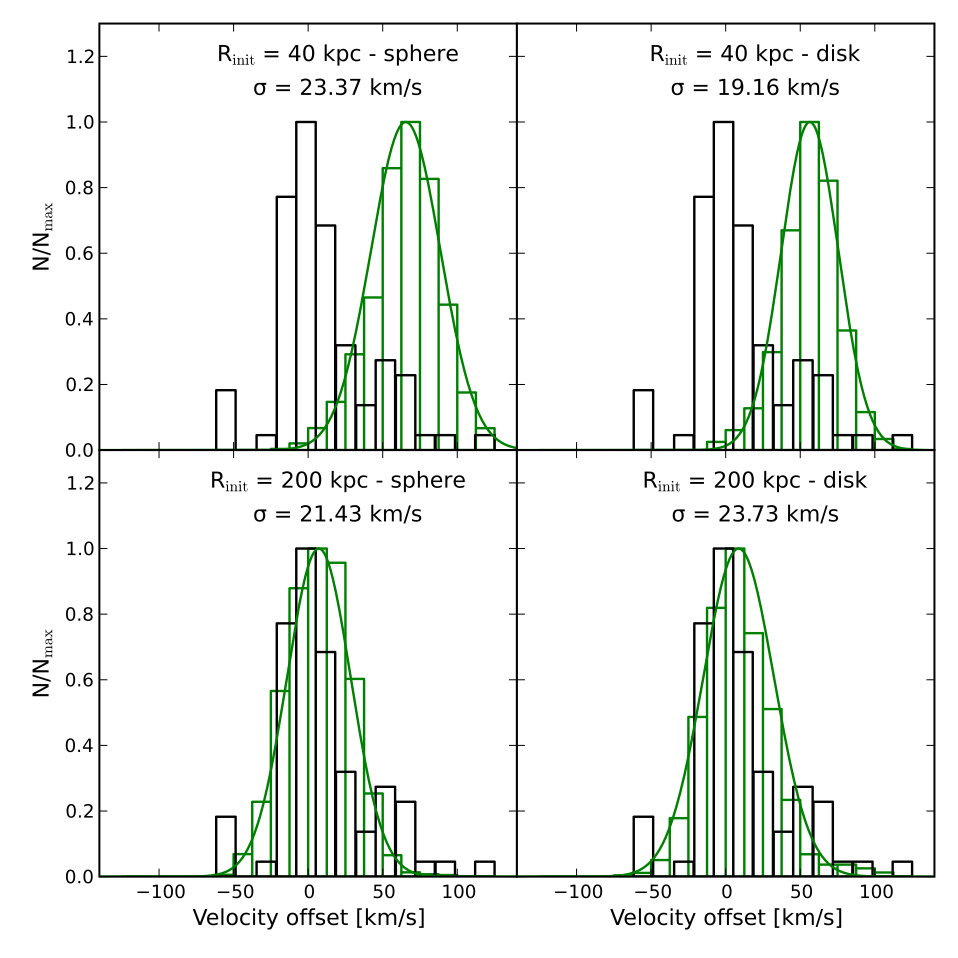}
\caption{
Velocity distribution of stream particles in four different $N$-body simulations of the models that successfully 
reproduce the positional data as well as the distance and velocity gradients along the stream. 
The black histogram is given by the observations \citep{Ibata2004} and the green by our 4 simulations.
The smooth green curves are the Gaussian fits.
The models with a dark-matter-poor satellite that use the initial phase-coordinates of equation \ref{eq:orbit_fardal} are 
in the top panels whereas our first-infall models, which use a dark-matter-rich satellite 
with the initial orbital parameters given by the equation \ref{eq:orbit_infall}, are shown in the bottom panels. 
The left column shows the case of an spherical satellite and 
the right column the case of a disk satellite. 
All the models are roughly consistent with the velocity dispersion derived 
from the observed GSS kinematics \citep{Ibata2004}. The
 cosmologically-motivated first infall model (lower panels) has the least offset w.r.t. the observations.
}
\label{fig:histovel}
\end{figure}

We show in Fig \ref{fig:histovel} the distribution of velocity offsets calculated for both the 
empirical modeling of the orbit with $R_{\mathrm{init}} = 40\,\kpc$ (first row), discussed in Section \ref{sec:empirical} and for our first-infall models 
with $R_{\mathrm{init}} = 200\, \kpc$ (bottom row). In each row, the left and right panels correspond 
respectively to the result obtained using a spherical and a disk satellite. In each panel, 
the green histogram is the result calculated from stellar particles in our simulated stream and 
the black histogram is the distribution from the observations \citep{Ibata2004}. We fit a gaussian distribution 
to the histogram to estimate the dispersion in the stream, similar to the procedure used by \citet{Ibata2004}.
They estimate the velocity dispersion in the stream to be $\sigma = 11 \pm 3 \kms$. 
We find that all  models tend to overestimate the dispersion in the stream. However, 
the best match between the observed and estimated distributions is obtained for our cosmologically-motivated 
first-infall scenario with a spherical Plummer satellite.
The small apocentre of the satellite trajectory in the empirical models, studied in Section \ref{sec:empirical}, is the reason for the systematic underestimation of the velocities.

\subsection{Dark-matter-poor versus dark-matter-rich progenitor satellite}

Our best fit model favours a dark-matter rich spheroidal dwarf galaxy as the progenitor of the GSS. 
However, one might argue that dark matter halo would play a marginal r\^ole in the
formation of the GSS, as most of it is stripped off the satellite long before it passes through M31.
In this section we shall use our N-body simulations to study this question.

We quantify the mass-loss
experienced by the satellite in our first-infall model. In our simulations, the dark halo is sampled with $N_h = 183333$ 
particles, yielding a mass resolution of $m_h = 1.68\,\times\,10^5\msolar$ for 
dark matter particles. The satellite starts initially at $r\sim 200\,\kpc$s 
and is followed up to $r\sim 40\,\kpc$s.

\begin{figure}
\centering
\includegraphics[width=\linewidth]{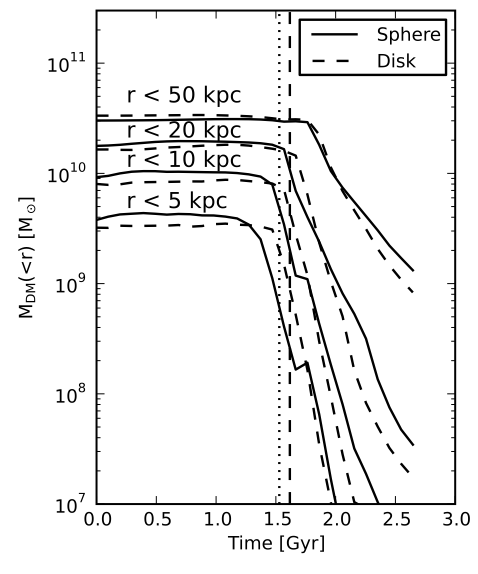}
\caption{Evolution of the enclosed dark matter mass $\mathrm{M_{DM}(<r)}$ 
of the satellite as a function of time for different radii $r$ in 
our first-infall $N$-body models. The solid and dashed line correspond
to a spherical or a disk progenitor respectively. The vertical dotted 
line indicate the time when the satellite is at $r \sim 40$ kpc, which is 
the starting point for various phenomenological models, discussed in Section \ref{sec:empirical} . The vertical dashed 
line corresponds to the pericentric passage. Due to the highly eccentric orbit required 
to reproduce the GSS, in both models the satellite is able to retain a significant fraction of its dark matter 
halo as it falls towards M31 from $200$ to $40$ kpc.
}
\label{fig:mass_loss}
\end{figure}

To obtain an estimate of the mass-loss from the satellite, we compute 
the dark matter mass $M_i(<r_i)$ that encloses a 
fixed radius $r_i$ as a function of time. The evolution for 
$r_i = 1$, $2$, $5$, $10$, $20$ and $50\,\kpc$s is plotted in 
Fig.~\ref{fig:mass_loss}. The pericentre occurs for $t/t_0\sim0.85$ 
where $t_0$ is defined as the time at which the satellite reaches $r\sim\,40\kpc$s. The halo is largely unaffected up to $t/t_0\sim 0.6$
at which point it starts to significantly deform due to the M31 tidal field.
Nevertheless, the satellite is able to retain a large portion of its mass 
up to $t = t_0$ which is when it reached at a 40 kpcs of M31. 
We conclude that dark matter halo can 
still contribute to a large fraction of the satellite mass up to the formation of GSS.  The orbital history of the satellite, 
reconstructed from simple numerical integration might give a contradictory result. However such integrations could be oversimplified and a proper 
N-body simulations, done here, is necessary to fully model the mass loss for the
unusual cases of highly eccentric orbits.

\section{The warp of M31 disk}
\begin{figure}
\centering
\includegraphics[width=\linewidth]{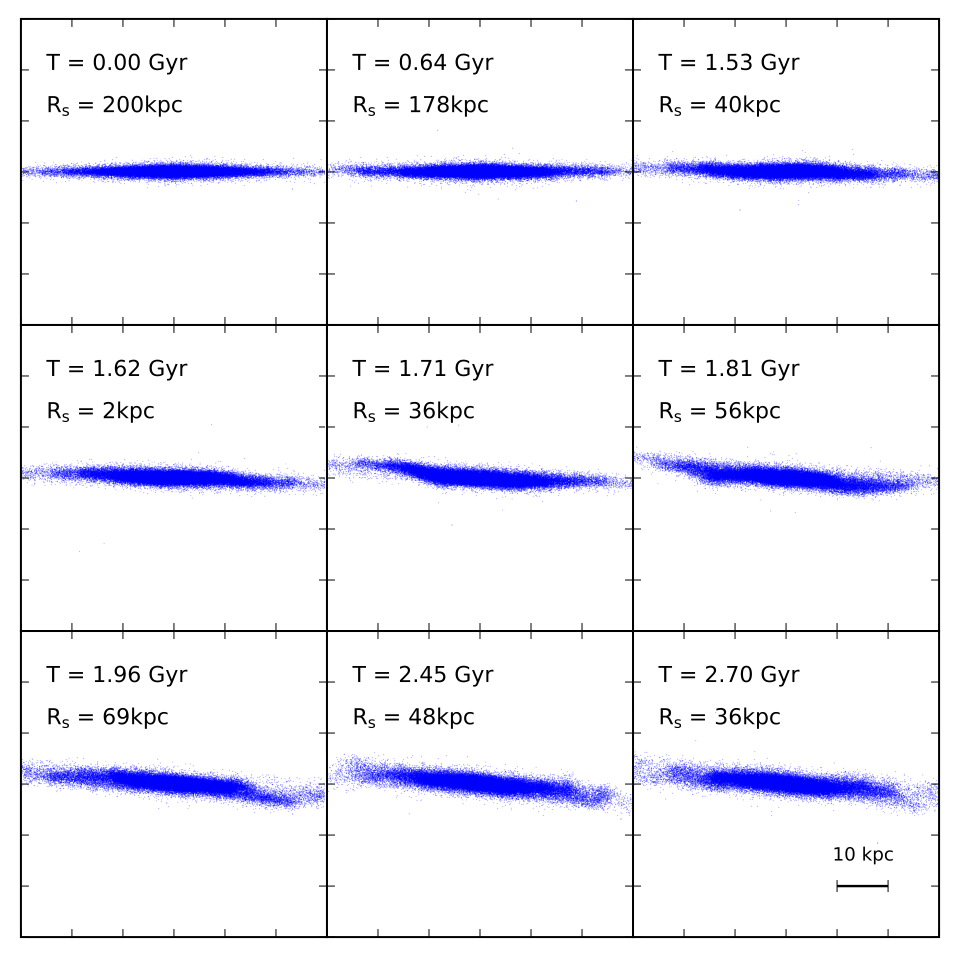}
\caption
{
Edge-on view of the M31 disk at different times in our first-infall model with a spherical infalling satellite.
The disk is visibly perturbed by the passage of the satellite and exhibits a warp-like structure 
at later times. The time after the start of the simulation and the orbital radius $\mathrm{R_s}$, which is the distance of the centre of mass of the satellite to M31,  is indicated in each frame.
}
\label{fig:warp of m31}
\end{figure}

The presence of a warp in the neutral hydrogen disk of M31 has been known for some times
\citep{baade1963,roberts1966,newton1977,whitehurstetal1978,innanen1982, Ferguson2002,Richardson2008}. 
Indeed, warps seem to be a common feature of many galaxies and it has been shown that
of the order of half of all galactic HI disks are
measurably warped, as is the disk of the Milky Way \citep{bosma1978}.  \citet{briggs1990} studied the warps of a sample of 12 galaxies in
details and inferred several general laws that govern the phenomenology of warps. In the years that followed
Briggs' work, many more catalogues of warp galaxies 
have been developed (see {\it e.g.} \citet{reshetnikov1999,sanchez2003}).
The fact that stellar warps usually follow the same warped surface 
as do the gaseous ones ({\it e.g.} \citet{cox1996}), is a strong evidence that warps are
principally a gravitational phenomenon.

The origin of warps remains unclear but 
numerous theories based on the interaction between the disk and the halo, or 
the cosmic infall and tidal effects, or nonlinear back-reaction from the spiral arms, or modified gravity 
have been proposed (e.g. see 
\citet{ostriker1989,quinn1992,binney1992,nelson1995,masset1997,jiang1999,
brada2000,lopez2002,sanchez2006,shen2006,weinberg2006}).

\begin{figure*}
\centering
\includegraphics[width=\linewidth]{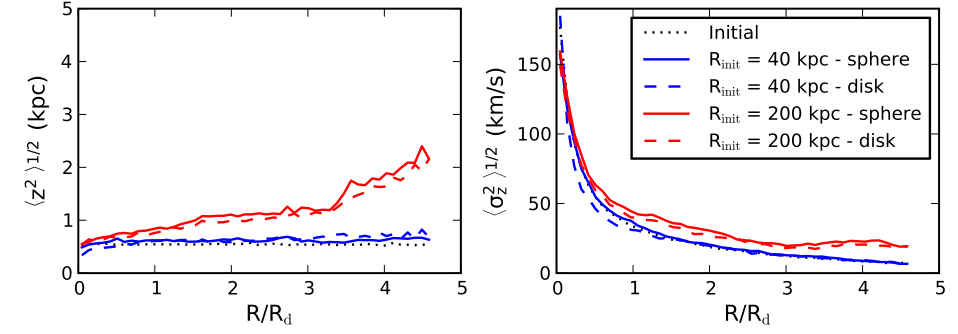}
\caption{
Heating of the disk of M31 due to the perturbations from the 
satellite in the $N$-body models that we studied in this work.
The results are shown for both disky and spherical progenitors and also for phenomenological models with a dark-matter-poor satellite ($R_{\rm init}=40$ kpcs), studied in Section \ref{sec:empirical}, and our first infall models with a dark-matter-rich satellite ($R_{\rm init}=200$ kpcs), studied in Section \ref{sec:infall}. 
Mean vertical height $\left<z^2\right>^{1/2}$ and the  
vertical velocity dispersion $\left<\sigma_z^2\right>^{1/2}$  as functions of cylindrical radius in units of the scale 
radius $R_d=5.4$ kpc  (see Table \ref{table:m31_model}) of the disk are plotted. In both of our first-infall models, the passage of the 
satellite significantly disturbs the disk because the progenitor retains a large 
fraction of its dark matter mass. The panels show that the outer regions 
of the disk ($\mathrm{R>2R_d}$) are heated and become thicker with respect to the inner parts, in our cosmologically-motivated first-infall model. 
}
\label{fig:diskheat}
\end{figure*}

In this work, we study the formation of the warp of M31 as a result of the infall of
the satellite progenitor of the giant stream. 
Up to now, we have only studied the tidal effect of M31 on the infalling satellite. However, as our satellite is massive, the disk of M31
could also become heated and perturbed during this infall. A few snapshots of the  evolution of the disk of M31 is shown from the beginning of the simulation 
 to the end in Fig.~\ref{fig:warp of m31}. The figure clearly shows that the disc
becomes tilted, heated and warped as the infalling satellite approaches and goes through M31.

To study these effects quantitatively, we first calculate 
the vertical height profile of the disk,  $<z^2>^{1/2}$, as a function of the cylindrical radius $R$
as shown in Fig.~\ref{fig:diskheat}. 
For comparison, we also plot 
the same profile calculated for the empirical models of GSS formation which rely on a dark-matter-poor satellite (discussed in Section \ref{sec:empirical}).
We find that, in our first-infall model, after the passage of the GSS progenitor,  
the disk thickness increases, as shown in Fig.~\ref{fig:diskheat}, and the scale height reaches about 2 kpcs, in good agreement with the observations that give an average scale height of  $2.8 \pm 0.6$ kpc for the thick
disk of M31 \citep{collins2011}. Although the satellite is dark-matter-rich and massive, its rapid passage through M31 guarantees 
that the disk of M31 is not destroyed or heated to extreme.  It has previously been suggested that minor mergers could be at the origin of the thick disks of 
galaxies \citep{purcell2010} and in our work we clearly see that
the infall of the progenitor of the GSS could be partially responsible for the thick disk of M31.

\begin{figure}
\centering
\includegraphics[width=\linewidth]{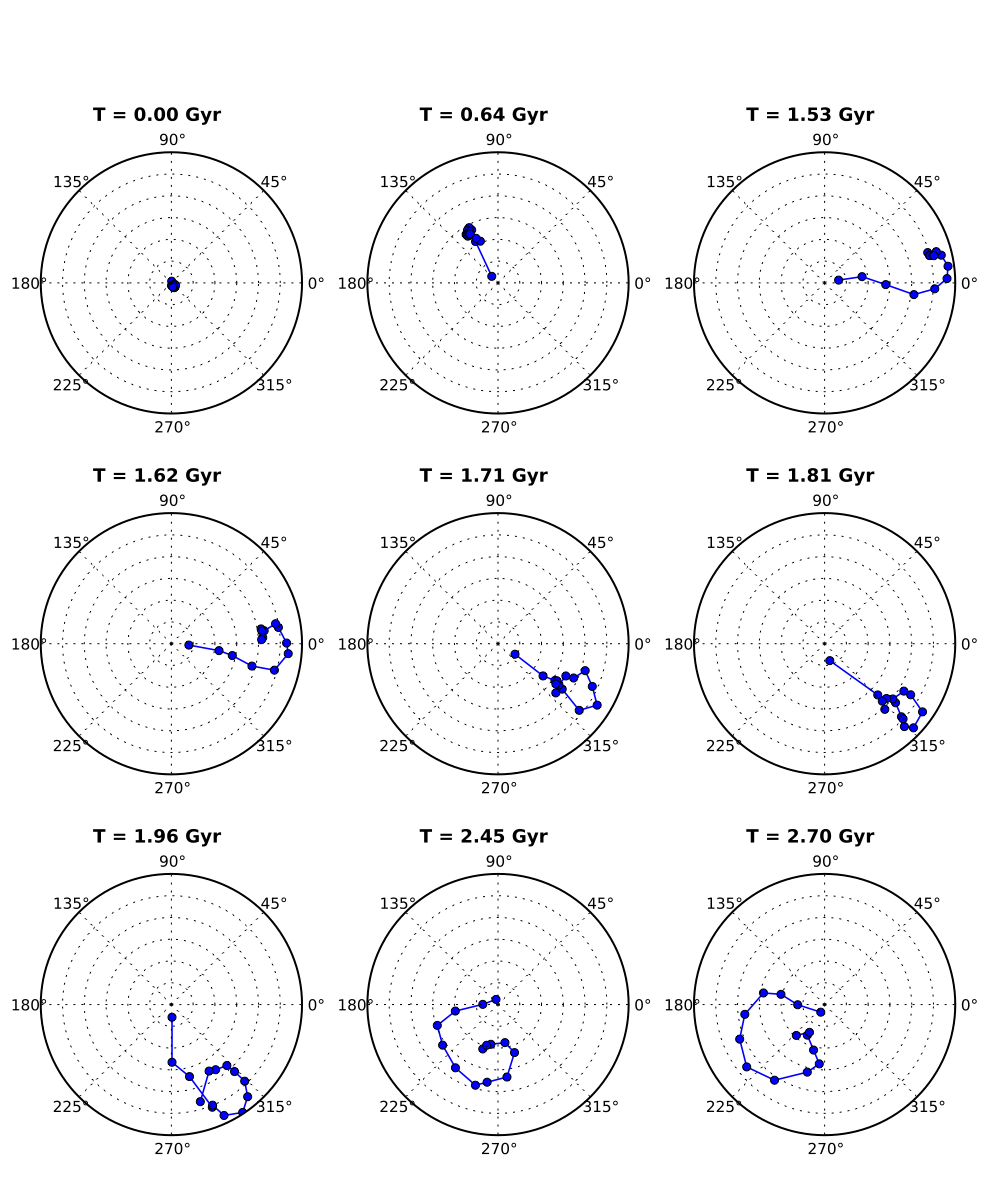}
\caption
{
The diagram of the line of nodes showing the time-evolution of the warp of M31 in our first-infall model, discussed in Section \ref{sec:infall}. 
The snapshots are shown from the start of the simulation (T=0 Gyr) to the last time-step (T=2.7 Gyr) which corresponds to the present time.
To make this diagram, the disk of M31 is divided into concentric annuli of width of about 1 kpc starting at 3 kpcs from the centre of M31 disk. The radial coordinate
 is the warp or inclination angle that an annulus of the disk  makes with the inner disk plane, shown at 
 one degree intervals.  The azimuthal coordinate gives the azimuth of the line of nodes. 
 The solid points are plotted for the radially-ordered annuli at 1 kpc intervals, apart from the first central point which 
 is averaged over 3 kpcs annulus. At the start of our simulation, there is no warp, and all points crowd at 
 the centre of the diagram. After the passage of the satellite through M31, the line of nodes first form a straight line 
 and then form a spiral when the differential rotation sets in  \citep{briggs1990}. The warp rotates clockwise and has a winding period of about 3 Gyrs.
}
\label{fig:tip-lon}
\end{figure}

The passage of the satellite through M31 also causes the disk of the galaxy to warp.
In general, the integral-sign warps can be viewed as the 
$m=1$ or s-wave perturbations that are excited in the disk 
by various sources \citep{hunter1969} and in our case by the passage of the satellite.
Warps are characterised by their lines of nodes and inclination angles \citep{briggs1990}. 
The diagram of  the line of nodes is an unusual polar plot of the angle made by the line of nodes and the latitude of the 
concentric rings into which the disk of the galaxy is divided, for the purpose of the study of the warps.
The line of nodes for our model develops into a spiral, as shown in Fig.~\ref{fig:tip-lon},  due to the differential rotation 
of the disk,  with a twist and a winding period of about 3 Gyrs which are all generic characteristics of galactic warps \citep{briggs1990,shen2006}. 
The observations find that the extended disk of M31 is about 30 kpcs and its HI warp starts at around 16 kpcs \citep{newton1977,henderson1979,brinks1984,chemin2009} and the scale-height of the 
gas layer reaches a maximum value of about 2 kpcs. These values agree reasonably 
with our results, although we find that the scale height starts increasing at
smaller radii for a stellar warp. 

\begin{figure}
\centering
\includegraphics[width=\linewidth]{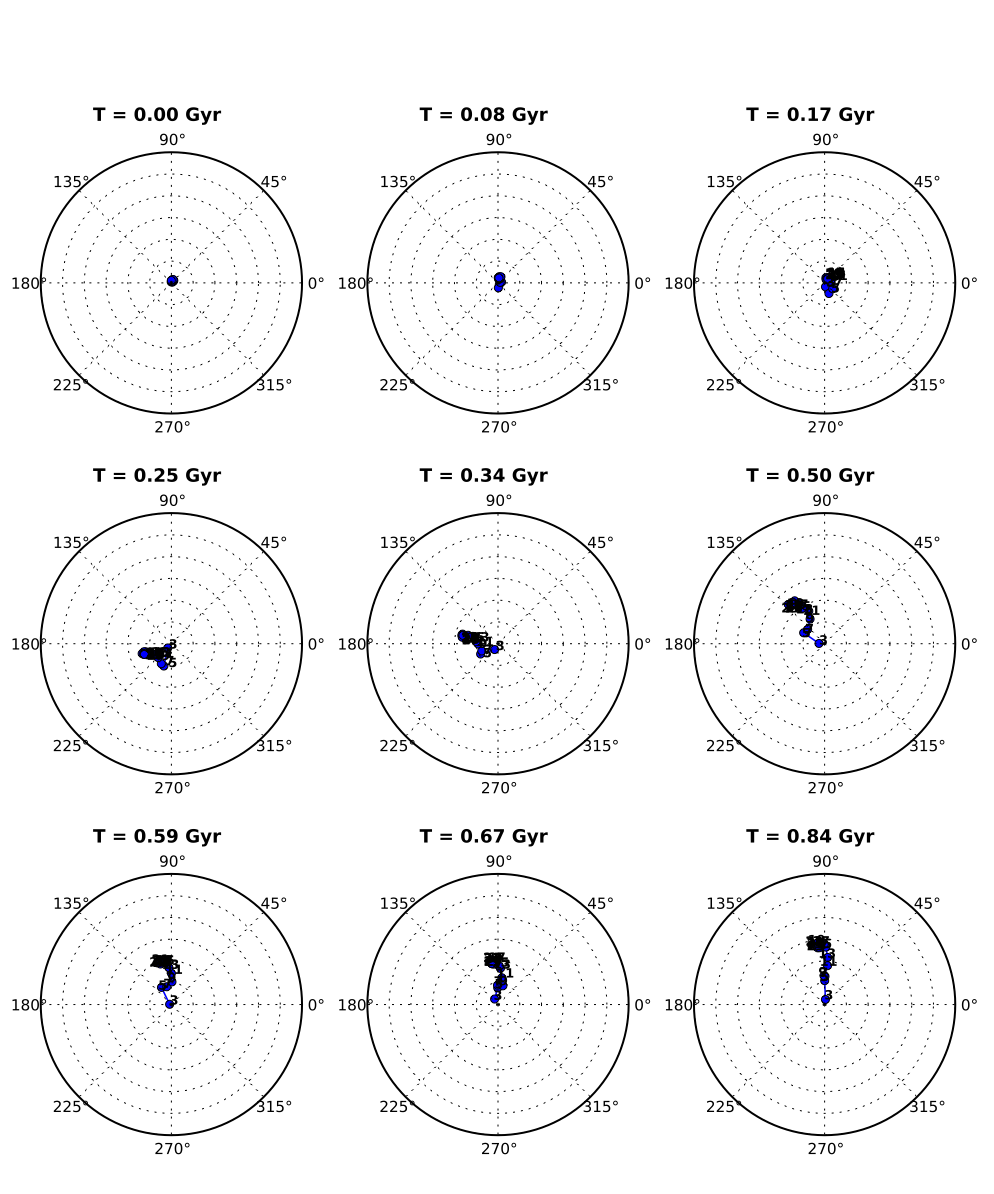}
\caption{
The figure is the same as Fig.~\ref{fig:tip-lon} but here is made for the phenomenological models (see Section \ref{sec:empirical}), with a spherical Plummer satellite.
The snapshots are shown for this model  from the start of the simulation (T=0 Gyr) to the last time-step (T=0.84 Gyr) which corresponds to the present time.
The satellite is dark-matter poor and falls from a short distance of 40 kpcs onto M31 and, as expected, only weakly perturbs the disk but cannot cause it to warp .
}
\label{fig:tip-lon-fardal}
\end{figure}

To demonstrate that the warps are indeed due to the passage of our massive satellite, we also 
make a similar plot of the line of nodes, in Fig.~\ref{fig:tip-lon-fardal}, for  the empirical model 
that we studied in Section \ref{sec:empirical}. Although perturbed and slightly heated, the disk is not 
warped in these models, which is expected since the satellite 
is dark-matter poor and starts its journey from a short distance of 40 kpcs from M31.

\section{Conclusion}
\label{sec:conclusion}

A wide range of observational data has progressively become available for the Andromeda galaxy. 
The disk of Andromeda is not flat but is distorted and warped. Its
 outskirts also seem drastically perturbed and 
a giant stellar stream, extending over tens of kiloparsecs, flows directly onto the centre of Andromeda.
The galaxy has about 30 satellites, observed so far, many of which seem to be corotating on a thin plane. 
These features have often been studied as unrelated events. Here we have aimed at providing 
a unique scenario that would fit these puzzling aspects of M31.   
We have shown that the accretion of a dark-matter-rich dwarf spheroidal provides a common origin for the giant southern 
stream and the warp of M31, and a hint for the origin of the thin plane of its satellites.

In our cosmologically-motivated model the trajectory  of the progenitor satellite lies on the same thin plane that presently contains many of M31 satellites
and
separates from the Hubble expansion at about 3 Gyrs ago and is accreted from its turnaround radius, of about 200 kpcs, into M31. It is disrupted as it orbits in 
the potential well of the galaxy and consequently forms the giant stream and in return heats and warps the disk of M31.
The position of the GSS 
and the two shelves are reproduced by our  full N-body simulations which uses a live M31.
The observed mass of the GSS obtained from its luminosity, is also predicted by our model,
which is in particular favoured by the kinematic data.
A prediction of our model is the actual position of the remnant of the progenitor satellite 
which should be found behind the north-eastern shelf.

As the satellite is dark-matter rich its infall perturbs the disk of M31. 
The thickness of the disk of M31 increases by a few kpcs and we have also shown
that the lines of node clearly indicate the presence of a warp with an angle going to about $6^\degree$, which agrees with the observations.

The stringent constraints set by a full range of observations on the initial conditions strongly suggest that the satellites of M31, which presently corotate on the 
same thin plane as our progenitor dwarf, could have similarly been accreted onto M31 along an intergalactic filament, which is yet to be identified by the observations.  The orbit of the progenitor satellite lies very close to the direction of M31-M33, as shown in Fig.~\ref{fig:thinplane-mw} and Fig.~\ref{fig:thinplane-m31}, which could 
be along an intergalactic filament  \citep{wolfe2013}, yet to be confirmed by observations.
Although not included, it is plausible that the gas which could have been contained within our massive satellite would also be disrupted during its passage through M31 and could contribute to the puzzling HI filament that joins M31 and M33 \citep{wolfe2013} which 
 also has been found to correlate, at least partially, with the giant southern stream of M31 \citep{lewis2013}.

\noindent
{\it\bf  Acknowledgments:}
We thank Wyn Evans, Francois Hammer, Rodrigo Ibata, Youjun Lu, Jim Peebles, Robyn Sanderson, Scott Tremaine, Qingjuan Yu for discussions.

\end{document}